\documentclass[aps,pre,longbibliography,preprint,superscriptaddress]{revtex4-1}
\pdfoutput=1

\usepackage[english]{babel}

\usepackage{graphicx}
\usepackage{amsmath}
\usepackage[colorlinks,linkcolor=blue,citecolor=blue,urlcolor=blue]{hyperref}

\begin{document}

\title{Collective Stochastic Coherence in Recurrent Neuronal Networks}

\author{Bel\'en Sancrist\'obal}
\thanks{Current address: Physics Department and Center for Neural Dynamics,
University of Ottawa, Ottawa, K1N 6N5 Canada; Mind, Brain Imaging and Neuroethics Research Unit, Royal Ottawa Healthcare,
Institute of Mental Health Research, Ottawa, K1Z 7K4 Canada}
\affiliation{Departament of Experimental and Health Sciences, Universitat Pompeu Fabra,
Barcelona Biomedical Research Park, 08003 Barcelona, Spain}

\author{Beatriz Rebollo}
\affiliation{Institut de Investigacions Biom\`ediques August
Pi i Sunyer, 08036 Barcelona, Spain}

\author{Pol Boada}
\affiliation{Institut de Investigacions Biom\`ediques August
Pi i Sunyer, 08036 Barcelona, Spain}

\author{Maria V. Sanchez-Vives}
\thanks{Corresponding authors (msanche3@clinic.ub.es, jordi.g.ojalvo@upf.edu)}
\affiliation{Institut de Investigacions Biom\`ediques August
Pi i Sunyer, 08036 Barcelona, Spain}
\affiliation{Instituci\'o Catalana de Recerca i Estudis Avan\c{c}ats
(ICREA), 08010 Barcelona, Spain}

\author{Jordi Garcia-Ojalvo}
\thanks{Corresponding authors (msanche3@clinic.ub.es, jordi.g.ojalvo@upf.edu)}
\affiliation{Departament of Experimental and Health Sciences, Universitat Pompeu Fabra,
Barcelona Biomedical Research Park, 08003 Barcelona, Spain}


\begin{abstract}
Recurrent networks of dynamic elements frequently exhibit emergent collective oscillations,
which can display substantial regularity even when the individual elements are
considerably noisy. How noise-induced dynamics at the local level
coexists with regular oscillations at the global level is still unclear.
Here we show that a combination of stochastic recurrence-based initiation with
deterministic refractoriness in
an excitable network can reconcile these two features, leading to maximum
collective coherence for an {intermediate noise level}.
We report this behavior in the slow oscillation regime exhibited
by {a cerebral cortex network under dynamical conditions resembling} slow-wave sleep and anaesthesia.
Computational analysis of a biologically realistic network model reveals
that an intermediate level of background noise leads
to quasi-regular dynamics. We verify this prediction experimentally
in cortical slices subject to varying amounts of extracellular potassium,
which modulates neuronal excitability and thus synaptic noise.
The model also predicts that this {effectively regular} state should exhibit
noise-induced memory of the spatial propagation profile of the collective oscillations, which is also verified
experimentally. Taken together, these results allow us to construe the enhanced regularity
observed experimentally in the brain as an instance of collective stochastic coherence.

\bigskip
\noindent
{\small
Final version published in {\em Nature Physics} vol. 12, 881-887 (2016),
DOI: \href{http://www.nature.com/nphys/journal/v12/n9/full/nphys3739.html}{10.1038/NPHYS3739}
}
\end{abstract}


\maketitle

Recurrent networks are directed graphs with
cyclic paths that can exhibit self-sustained collective dynamics. When the network nodes 
are threshold elements, a sufficiently large background
noise will render their activity stochastic. Yet, the collective behavior of the network
is frequently highly regular in time. This raises the question of how the stochastic nature of the network
elements coexists with the quasi-deterministic character of the collective dynamics.
While coupling has long been proposed as a regularizing mechanism for interacting sloppy
oscillators
\cite{Enright:1980fk,Garcia-Ojalvo:2004uq}, the situation is much less clear when
the individual elements are not intrinsic oscillators, but exhibit noise-driven pulsatile dynamics,
such as in excitable elements. A relevant instance of this situation is given by neuronal
networks.

Here we study the interplay between noise and collective dynamics in networks
of neurons from the cerebral
cortex operating in the state of slow oscillations, a dynamical regime that has been suggested as the default activity of the cortex \cite{Sanchez-Vives:2014fk}.
In this physiological state, typical of slow wave sleep and anesthesia \cite{Steriade1993b},
the membrane potential of cortical neurons alternates at frequencies
of the order of 1~Hz between the so-called
UP and DOWN states \cite{Stern1997,Shu2003}.
UP states are characterized by a depolarization
of the membrane voltage towards the spiking threshold and a sustained firing activity of the
neurons, similar to their dynamics during wakefulness. In contrast, in the DOWN states
neurons are mostly silent and exhibit a hyperpolarized membrane voltage.
The fact that UP and DOWN states exist spontaneously {\em in vitro} \cite{Sanchez-Vives2000,Compte:2008fk}, in the absence of external stimulation,
suggests that this dynamical regime is self-sustained, appearing locally without requiring
large-scale cortical interactions nor external inputs. In other words, the recurrent connectivity
between neurons is sufficient for the emergence of these slow oscillations \cite{Mattia:2012uq}.

We examine the recurrent network dynamics exhibited {\em in vitro} by
slices of the ferret cerebral cortex. In our experiments, noise is determined by the level of neuronal
excitability, which can be controlled by the extracellular potassium concentration in the medium.
In contrast to previous studies \cite{PhysRevLett.77.4098}, no external time-dependent signals
are applied to the system, which operates spontaneously in a regime very close to what is observed
{\em in vivo} \cite{Sanchez-Vives2000}.
The regularity of the
slow oscillations decreases when the brain comes out of deep anaesthesia \cite{Deco2009},
anticipating the loss of the slow oscillatory regime and the emergence of the sustained depolarized
state characteristic of wakefulness \cite{Steriade2001}.

Theoretical work has shown
that the sequence of UP and DOWN states can be highly irregular in the
presence of noisy inputs provided inhibition is decreased \cite{Parga2007}, for low AMPA conductances
of the connections between pyramidal neurons and inhibitory interneurons \cite{Bazhenov2002},
or more generally by changing the stability balance between metastable UP and DOWN attractors,
for instance through modulation of the fatigue- and adaptation-dependent inhibitory feedback
\cite{Mattia:2012uq}.
The nature of the more regular regime characteristic of the sleep state is, however, still under debate.
Is it a deterministic or a noise-driven state? Here
we explore the latter possibility in an isolated network of the cerebral cortex.
We observe that the cortical
network generates slow oscillations that exhibit maximal regularity for an {intermediate} amount
of background synaptic noise, in what can be construed as an instance of stochastic coherence. 

In generic (uncoupled) excitable systems, stochastic coherence arises from the fact that
the entry into the excited state is noise-driven, whereas the exit is basically controlled by
deterministic processes and is followed by a refractory period, after which no reinitiation
is possible for a certain time window under moderate noise \cite{Pikovsky1997,Lindner2004}.
In those conditions, increasing noise leads to more frequent initiations that eventually
pile up at the refractory period with strong regularity, while for larger noise refractoriness
breaks down and the dynamics becomes irregular again. Here we conjecture that UP/DOWN
oscillations exhibit this behavior, motivated by the realization that the neuronal networks
underlying the phenomenon exhibit the above-mentioned three main requirements of stochastic coherence
(stochastic initiation, deterministic termination, and refractoriness) at the collective level.

First, the initiation of the UP state is triggered by the spontaneous (noise-driven) simultaneous
firing of a few excitatory neurons.
Second, the termination of the UP state can be explained by an accumulation of
adaptation during the UP state, which can be expected to be mainly
deterministic \cite{Mattia:2012uq}. The transition from the UP to
the DOWN state can be accounted for by the activity-dependent adaptation of neurons caused
by the activation of potassium channels that reduce the sensitivity to synaptic inputs
\cite{Sanchez-Vives2000,Compte2003}. This is a cumulative processes driven by the high (and well-defined)
firing rate characteristic of the UP regime, and is thus essentially deterministic. Other cumulative mechanisms
for UP-state termination have been proposed, including synapse fatigue caused by depression
\cite{Holcman2006,Mejias2010}, but they are also mostly independent of noise at the network level.
Importantly, all these mechanisms require a recovery of the network excitability following the UP state,
leading to refractoriness \cite{Sanchez-Vives2000}.

We thus propose that stochastic coherence provides a mechanism by which the natural
excitability of the cortex determines the regularity of the UP/DOWN dynamics, by controlling the amount of background synaptic noise acting
upon the neurons. In that way,
an {intermediate level of excitability which coincides with physiological conditions
\cite{Amzica:2002uq}} would lead to maximally coherent slow oscillations. 
In that scenario, the variations in regularity that are characteristic of sleep-to-wake transitions
\cite{McCormick1989,Romcy-Pereira2009} would arise from regulated changes in cortical excitability.
In contrast with previous studies of stochastic coherence at the level of individual neurons
\cite{Pikovsky1997,Han1999,balenzuela2005role}, the stochastic
coherence reported here is a purely collective phenomenon, since both the initiation and the termination
of the UP states arise only at the network level (initiation resulting from recurrent activation, and
termination emerging from potassium channel-mediated adaptation, which only comes into effect
during network-driven, high-frequency UP state activity).

\subsection*{Collective stochastic coherence in a spiking network model}

\begin{figure}[htb]
\centerline{
\includegraphics[width=0.65\textwidth]{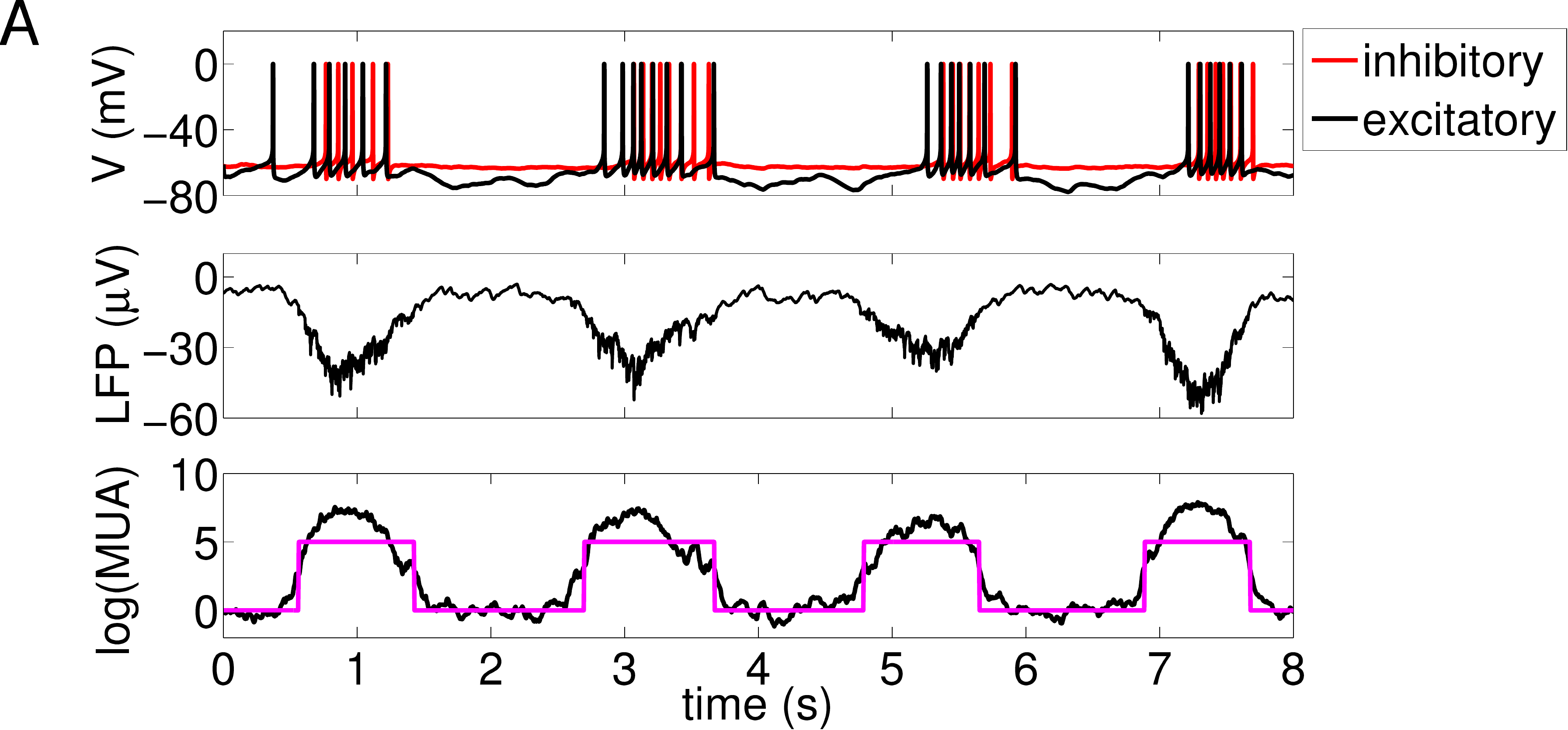}
}
\vskip2mm
\centerline{
\includegraphics[width=0.3\textwidth]{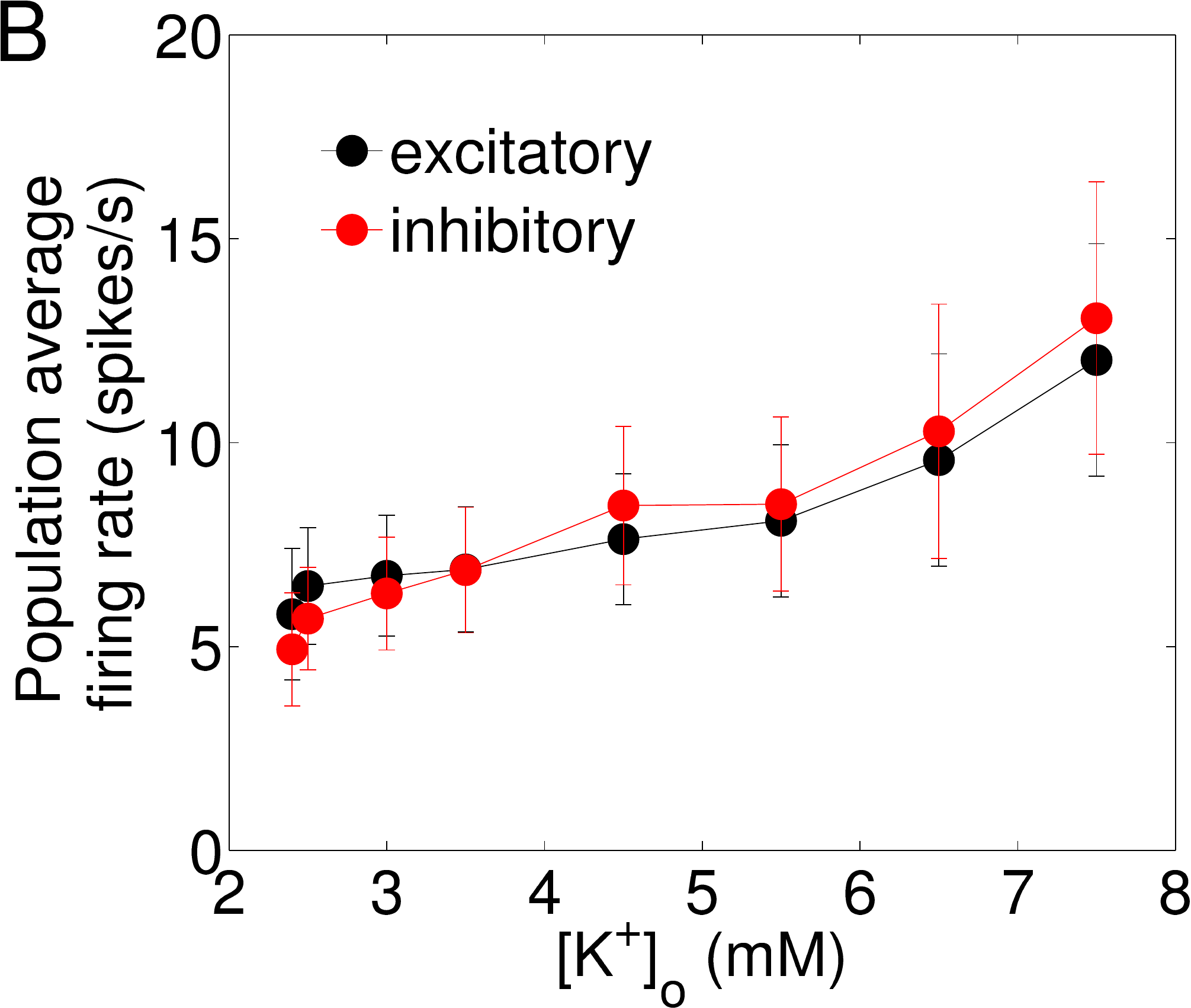}~~~
\includegraphics[width=0.3\textwidth]{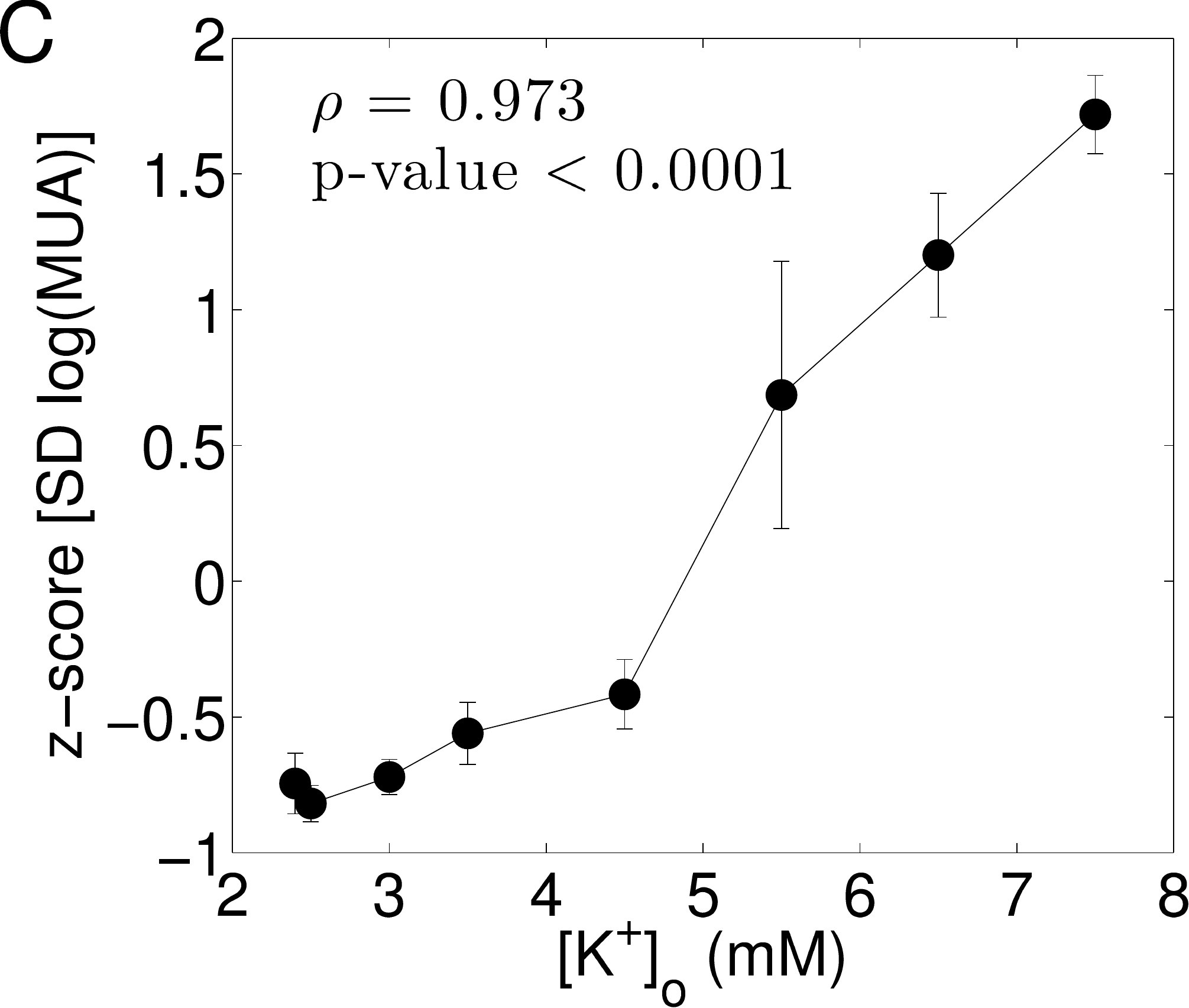}
}
\caption{UP/DOWN oscillations in a cortical network model. A, time traces of
two specific neurons in the network (top panel), including an excitatory neuron (black line)
and an inhibitory neuron (red line). The LFP and log(MUA) time traces of the full network
are shown in the middle and bottom panels, respectively. Magenta lines in the bottom panel indicate the boundaries of the UP states computed from the log(MUA) signal.
B, average firing rate of the excitatory and inhibitory subpopulations of the complete network
during the UP state for varying extracellular potassium concentration.
C, z-score of the log(MUA) standard deviation in the DOWN state for increasing extracellular
potassium concentration. Error bars in panels B and C represent the corrected sample standard deviation. Computations are made across 5 different simulations with distinct realizations of the connectivity matrix and external input. $\rho$ denotes the Pearson correlation coefficient.
\label{fig:moddyn}}
\end{figure}

Following Compte et al \cite{Compte2003}, we consider a network of excitatory and inhibitory neurons
described by conductance-based models (see Online Methods for a full description).
Network clustering enables recurrence of neuronal activity, through which randomly
occurring spikes lead to a cascade of neuronal firing events (UP state). This is shown in the top panel of
Fig.~\ref{fig:moddyn}A, which depicts the typical
dynamics of two neurons from the network, one excitatory and the other one inhibitory.
The UP state terminates mainly
via a potassium channel dependent on the intracellular sodium concentration,
$I_{\rm KNa}$, which is known to be expressed throughout the brain \cite{Bhattacharjee2005a}.
An increase in the sodium concentration due to the enhancement of the firing activity of the cell
during the UP state activates this adaptive current, which in turn renders the neuron insensitive to upcoming
presynaptic action potentials \cite{Wang2003}. This mechanism leads to a
hyperpolarization of the neurons 
that increases in amplitude and duration with the firing rate \cite{Sanchez-Vives:2000uq,Sanchez-Vives:2010fk},
and terminates the UP state \cite{Compte2003}.
To quantify the dynamics of the entire neuronal population we use the multiunit activity (MUA), whose value in logarithmic scale is estimated as the average spectral power of the local field potential (LFP) in a particular frequency band, relative to the total power \cite{Mattia:2002kx} (see Online Methods). The time evolution of the LFP (Fig.~\ref{fig:moddyn}A, {middle}) and of the log(MUA) (Fig.~\ref{fig:moddyn}A, bottom)
exhibit clear slow oscillations, which reflect the cyclic time course of the synaptic currents
flowing within the network and, at the same time, the firing activity of neurons.     

The UP states shown in Fig. \ref{fig:moddyn}A are driven by background synaptic activity
impinging stochastically on each neuron, coming
from their presynaptic neighbors. The main way in which this
background synaptic noise can vary is through changes in the excitability of the local network.
To modify the excitability (ignoring external inputs) and thus change
the background synaptic noise, we varied the resting membrane potential of
all neurons by acting upon the equilibrium potential of potassium, which
plays the largest role in establishing the neuron's resting potential.
This can be replicated experimentally by modifying the concentration
of potassium in the extracellular medium, [K$^+$]$_{\rm o}$. In fact, recordings performed in cats during transitions from slow-wave to REM sleep showed an enhancement of [K$^+$]$_{\rm o}$ \cite{Satoh:1979uq}, suggesting its influence in neuronal excitability and network dynamics. In our model we increased [K$^+$]$_{\rm o}$ from 2.4~mM to 7.5~mM,
examining its effect on the dynamics of the individual
neurons. The firing rate during the UP states, shown in
Fig.~\ref{fig:moddyn}B for both excitatory and inhibitory neurons, reveals that
the firing activity of both neuron types increases with [K$^+$]$_{\rm o}$.
Note that inhibition overcomes excitation  for high excitability, 
preventing runaway activity that would otherwise appear in this recurrently connected model network.

To establish the relationship between network excitability and the initiation of
UP states, we next quantify the activity of the network during the DOWN states (which is small but non-negligible).
This activity is estimated by the log(MUA), which is known to correlate with spiking activity \cite{Mattia:2002kx,Sanchez-Vives:2010fk}. Variability in this quantity during the DOWN states can thus be associated with background
synaptic noise. We compute
the standard deviation of the {log(MUA)} during the
DOWN states as a function of [K$^+$]$_{\rm o}$. The result, shown in Fig.~\ref{fig:moddyn}C,
reveals a clear increase of the standard deviation as the extracellular potassium level grows {(with a high linear correlation, $\rho$, and low p-value)}. We thus identify the excitability, controlled by [K$^+$]$_{\rm o}$, with the noise
acting upon the network. 

\begin{figure}[htb]
\centerline{
\includegraphics[width=0.58\textwidth]{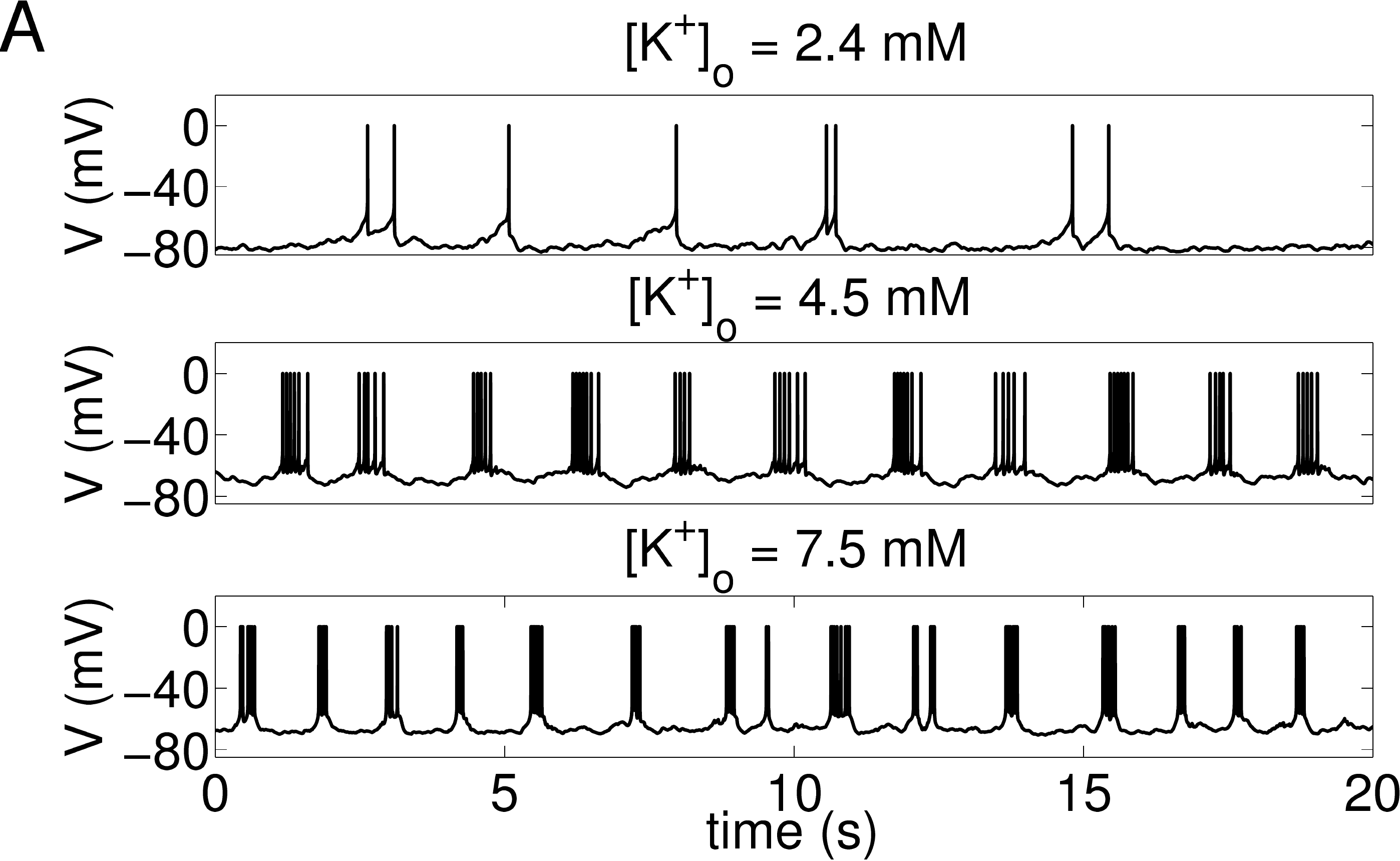}
}
\vskip2mm
\centerline{
\includegraphics[width=0.3\textwidth]{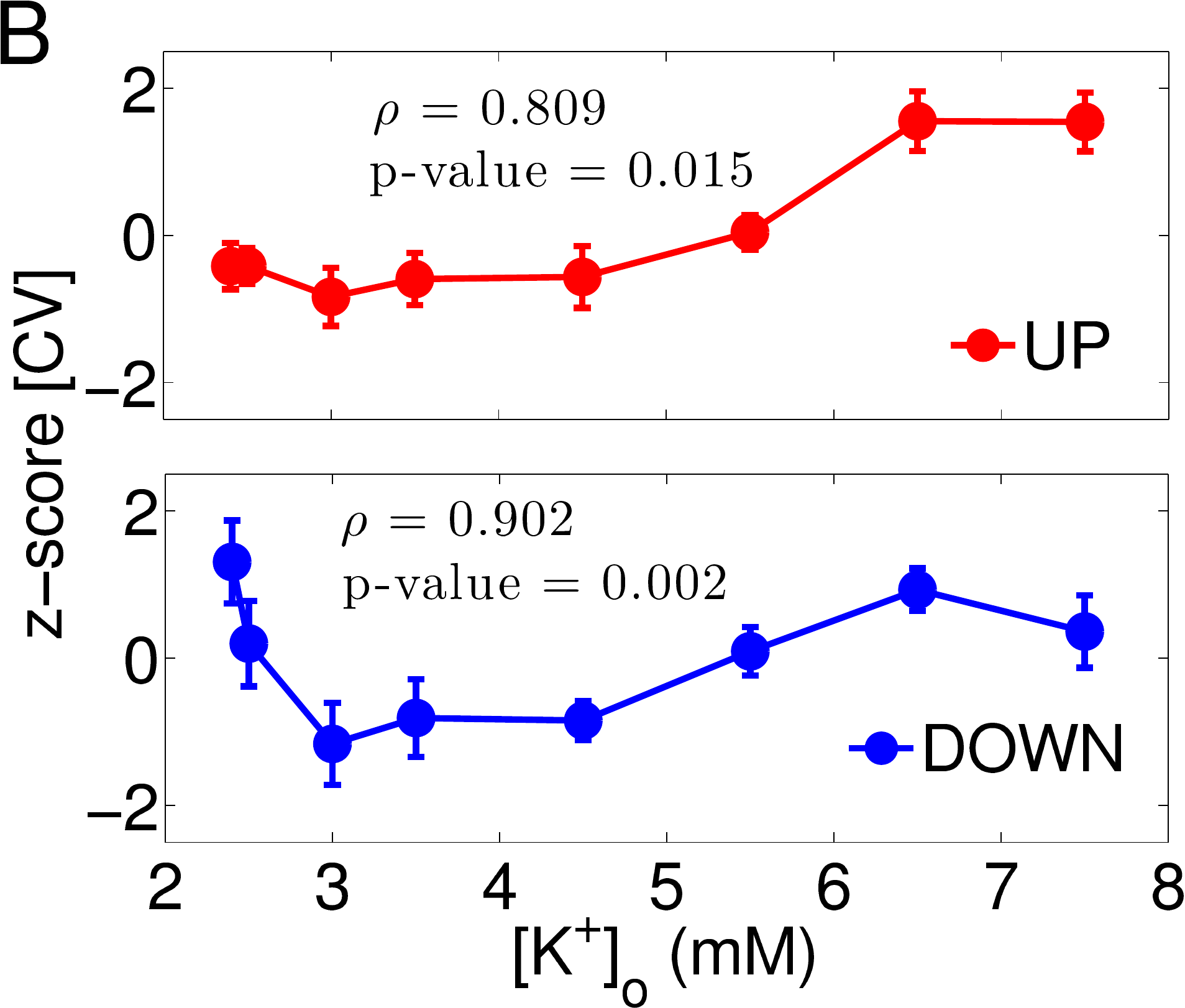}~~~
\includegraphics[width=0.3\textwidth]{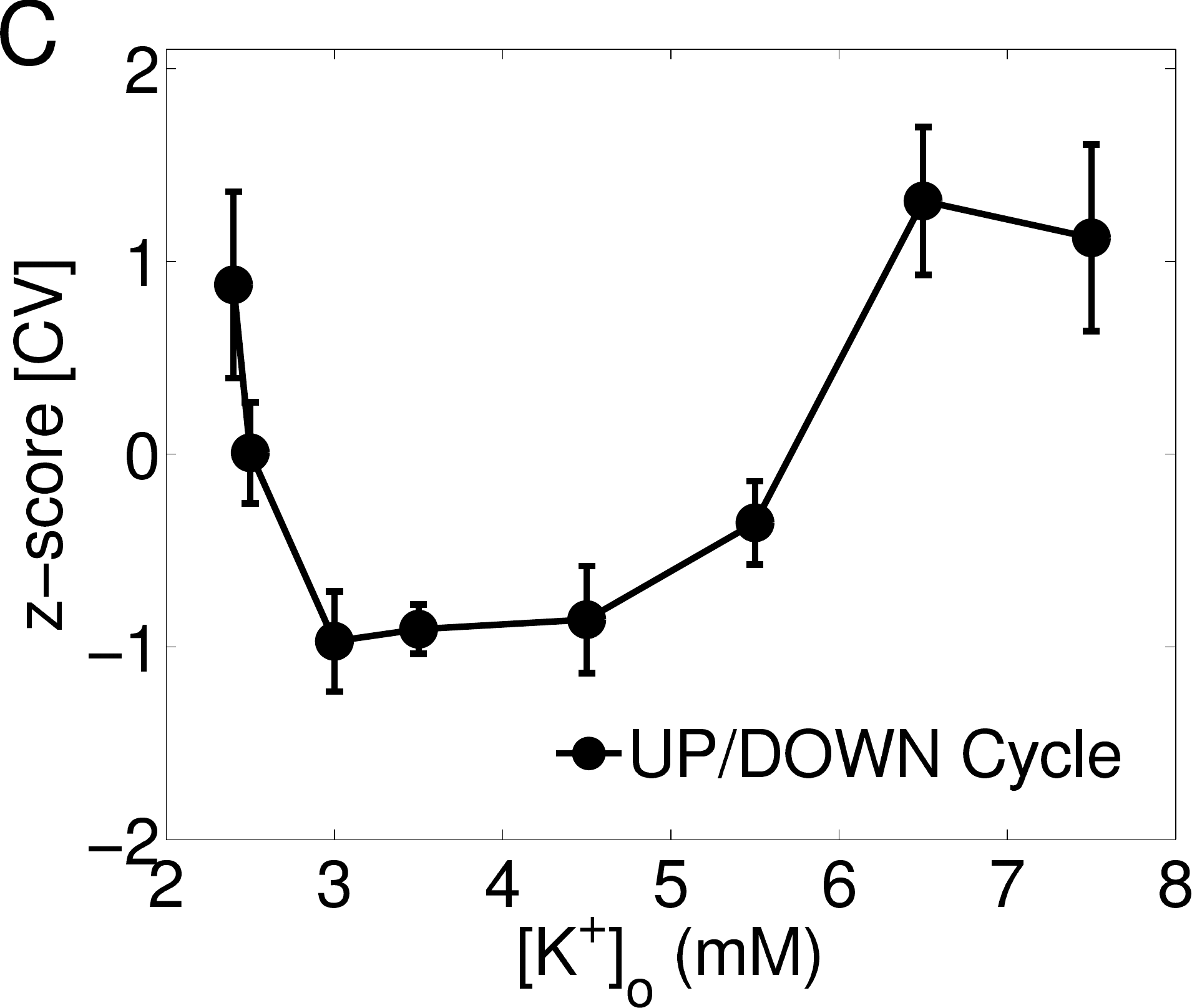}
}
\caption{Stochastic coherence in a cortical network model. A,
temporal evolution of the membrane potential of a representative excitatory neuron
of the network for three different values of the extracellular potassium concentration
[K$^+$]$_{\rm o}$. B, coefficient of variation (CV) of the duration of the UP
and DOWN events (red and blue, respectively). {The Pearson coefficient $\rho$ and the p-value quantify the correlation between each CV and the one shown in C, and the statistical significance of such correlation, respectively.}
C, CV of the duration of the full UP/DOWN cycle. Note that z-scores are shown in the y axes of panels B and C.
In those panels, error bars denote corrected sample standard deviation.
\label{fig:stcoh_mod}}
\end{figure}

The response of the network to an increase in excitability is shown in Fig.~\ref{fig:stcoh_mod}A.
The figure depicts the temporal evolution of the membrane potential of a representative neuron of the model network for three different
values of the extracellular potassium concentration. For low [K$^+$]$_{\rm o}$ (top panel) the excitability
level is small, and consequently the initiation of UP states is infrequent. As the excitability increases
(middle and bottom panels of Fig.~\ref{fig:stcoh_mod}A) the UP events become more frequent, due
to higher noise levels, and the firing rate during the UP state also increases (in agreement with
Fig.~\ref{fig:moddyn}B).

We next examine how the variability of the oscillatory dynamics depends on noise (excitability),
by computing the coefficient of variation of the durations of the UP and DOWN events.
As shown in Fig.~\ref{fig:stcoh_mod}B, for low excitability levels the variability of the UP
phase duration is basically constant, in agreement with our assumption that termination
is essentially deterministic unless noise is too large. In contrast, the variability of
the DOWN phase duration sharply decreases with noise for low noise levels,
growing again when noise dominates the dynamics. The CV of the full cycle duration (Fig.~\ref{fig:stcoh_mod}C)
follows closely that of the DOWN state, showing a clear minimum for intermediate noise (excitability), which is the main
hallmark of stochastic coherence. 
This behavior is not due to the concomitant increase in firing rate arising from growing excitability
(Fig.~\ref{fig:moddyn}B): as shown in the Supp. Information (Section S3),
increasing the external noise acting upon the neurons produces the same effect with essentially no variation
in the firing rate.

\subsection*{Evidence in cortical slices}

\begin{figure}[htbp]
\begin{center}
\begin{minipage}[c]{0.40\textwidth}
\mbox{}~~~~\includegraphics[width=0.9\textwidth]{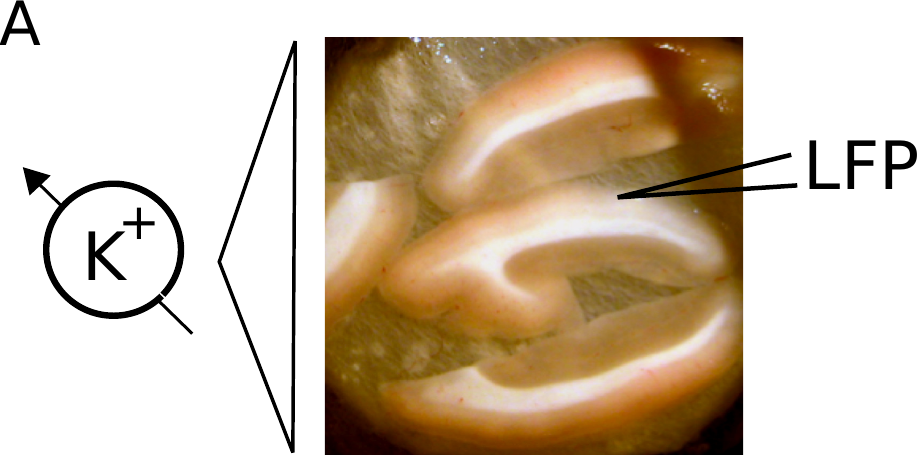}
\vskip1mm
\includegraphics[width=0.78\textwidth]{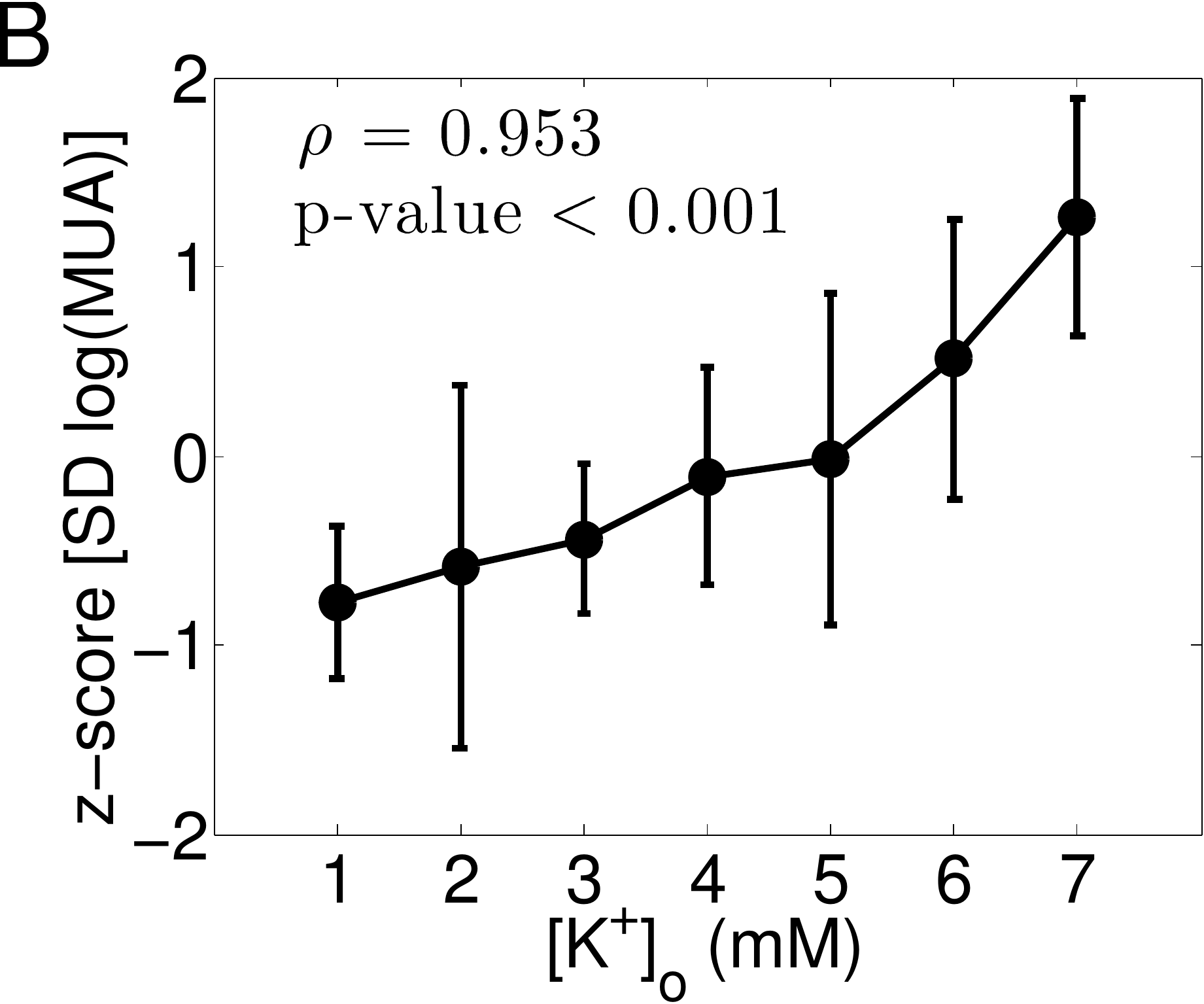}
\end{minipage}
~~
\begin{minipage}[c]{0.55\textwidth}
\centerline{
\includegraphics[width=\textwidth]{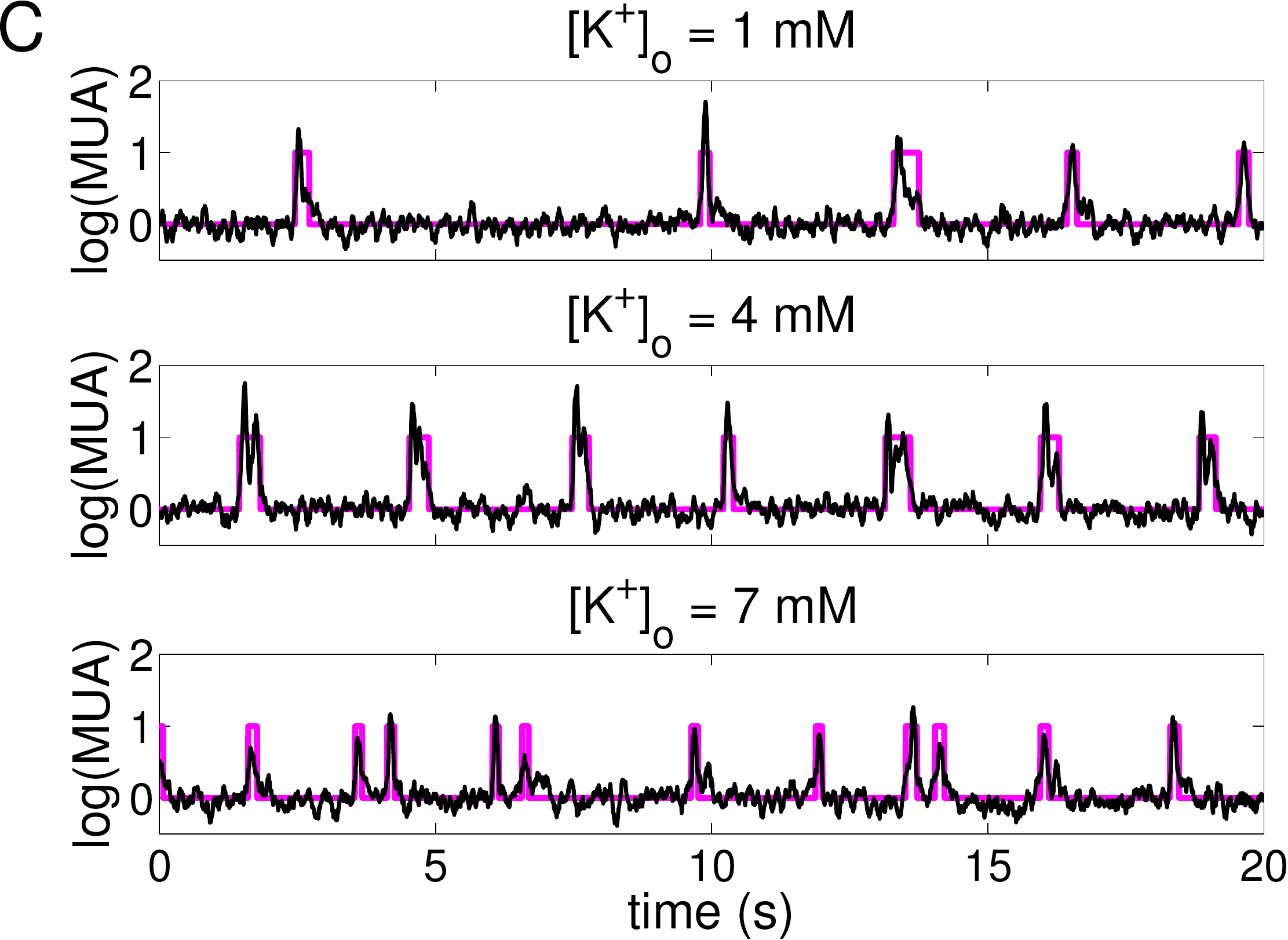}
}
\end{minipage}
\end{center}
\centerline{
\includegraphics[width=0.29\textwidth]{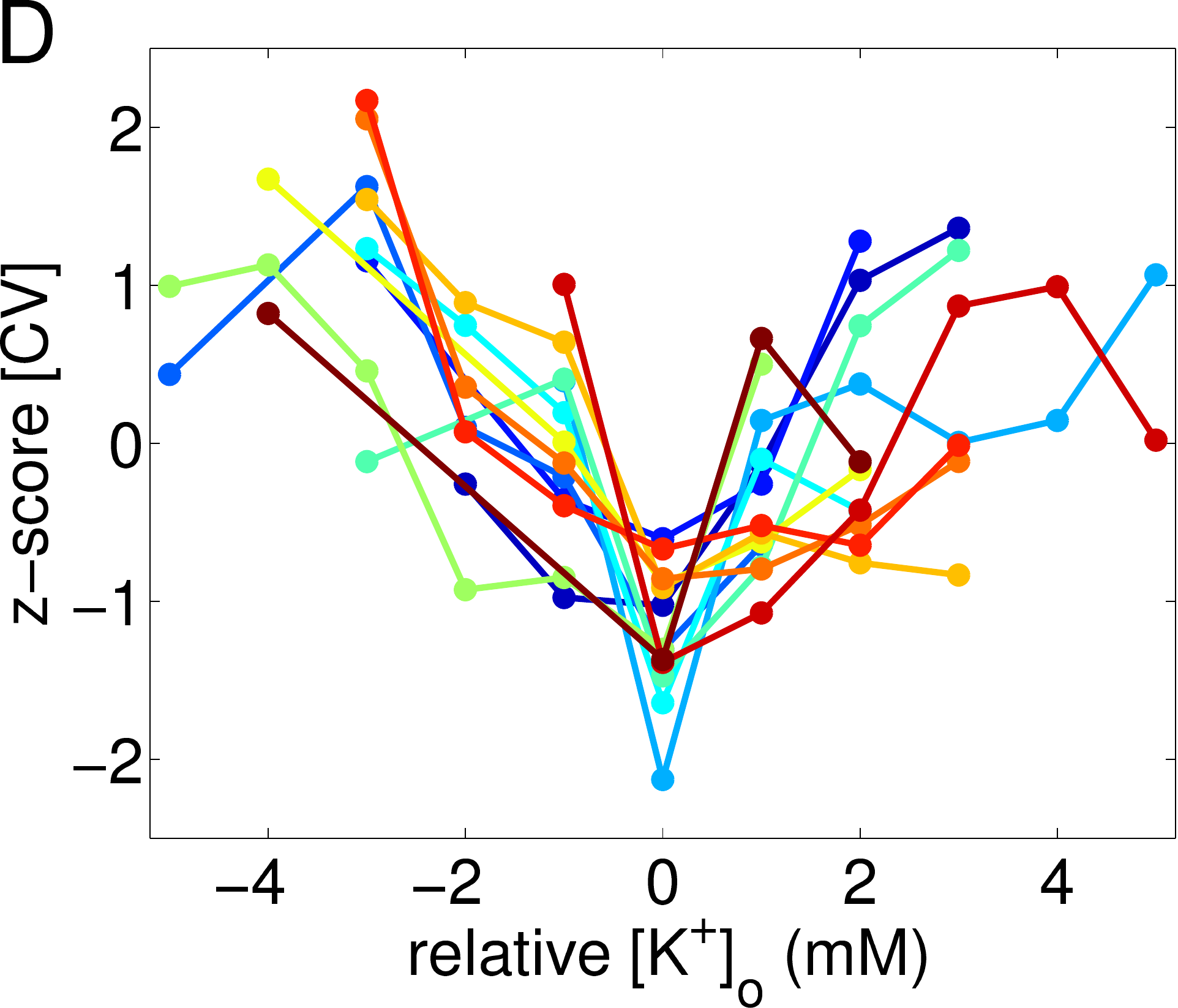}~~~
\includegraphics[width=0.29\textwidth]{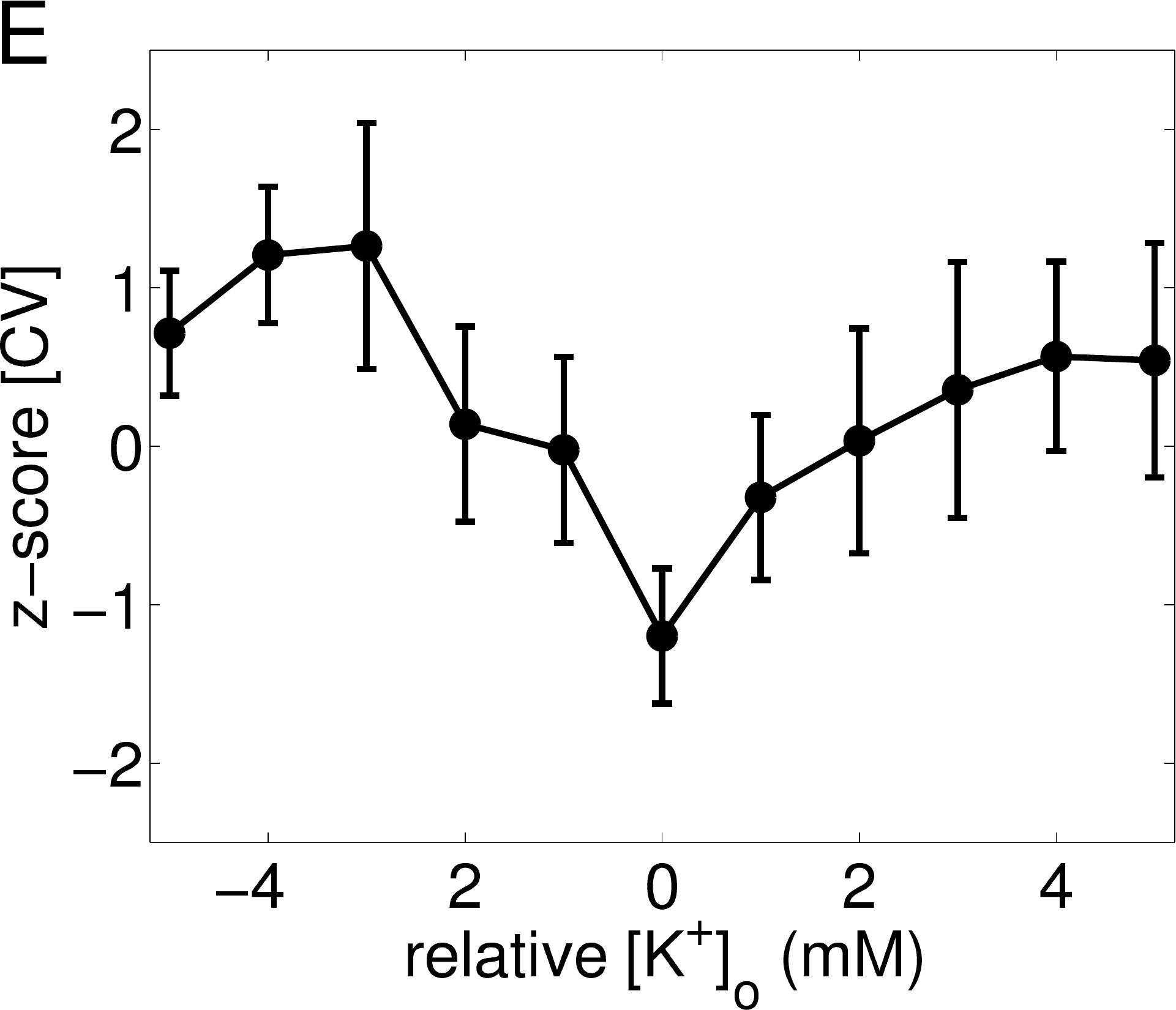}~~~
\includegraphics[width=0.29\textwidth]{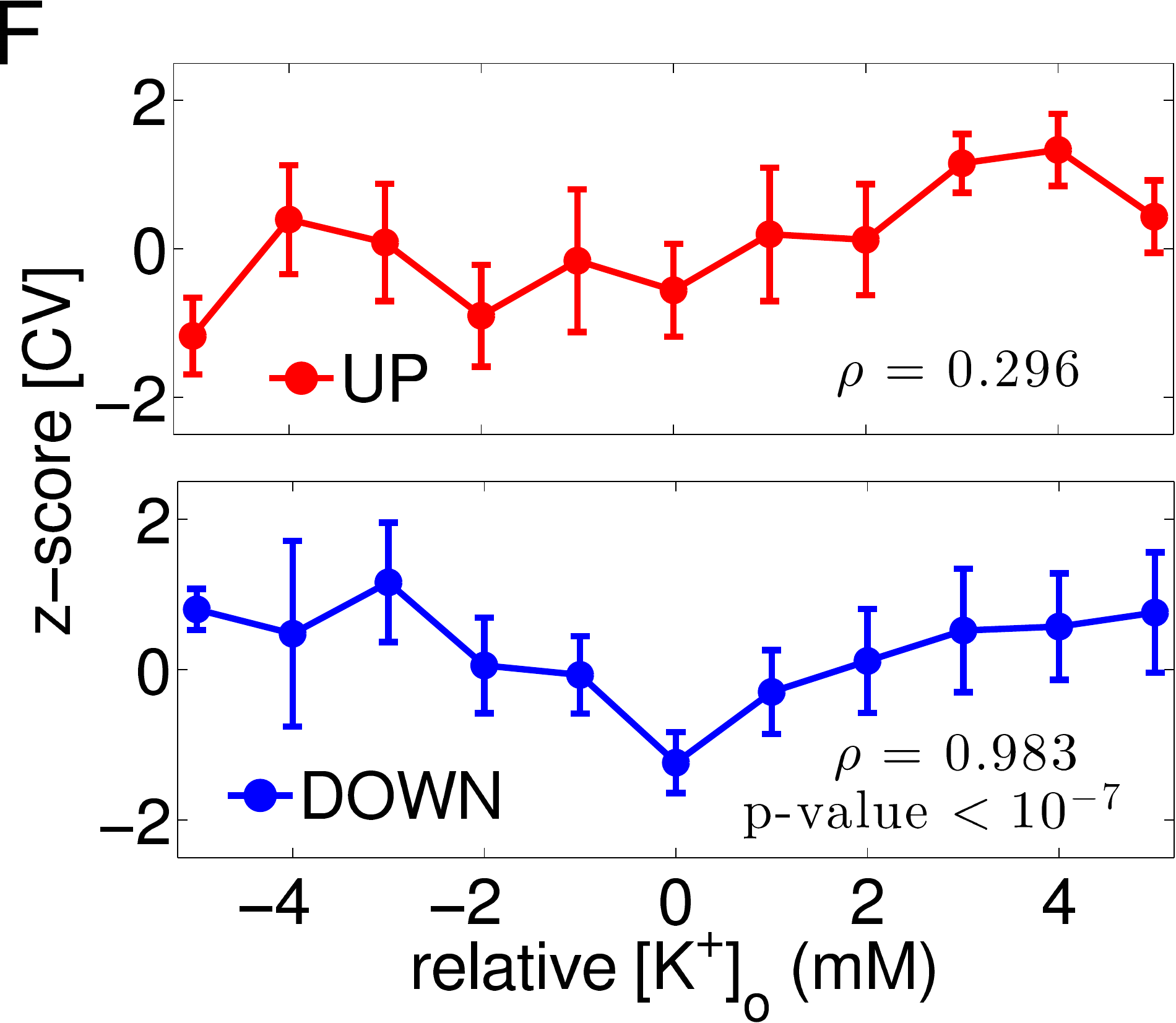}
}
\caption{Experimental evidence of stochastic coherence in the cortical tissue.
A, illustration of the experimental procedure: three visual cortex slices maintained in an interface recording chamber {\em in vitro}. B, effect of varying extracellular potassium
concentration [K$^+$]$_{\rm o}$ on the standard deviation of the log(MUA) during the DOWN
states. 
C, typical time traces of the {log(MUA)} for three different levels of [K$^+$]$_{\rm o}$. 
D, CV of the UP/DOWN cycle period versus [K$^+$]$_{\rm o}$ for all the different experimental trials ($N=13$ slices).
E, average CV of the UP/DOWN cycle period.
F, average CV of the individual durations of the UP and DOWN states versus
[K$^+$]$_{\rm o}$. To account for variability across experiments, in panels D-F the curves are aligned horizontally so that the
minimum CV of the UP/DOWN cycle corresponds to [K$^+$]$_{\rm o}=0$. Error bars in panels B, E, and F correspond to the corrected sample
standard deviation.
\label{fig:stcoh_exp}}
\end{figure}

To validate experimentally the prediction made by our computational model we turned to {\em in vitro}
cortical slices of visual cortex of ferrets, since UP/DOWN transitions have been observed in cortical slices under different
conditions \cite{Sanchez-Vives2000,Shu2003,Cossart2003}.
We studied the rhythmic activity patterns generated under various levels of synaptic noise by
varying the extracellular potassium concentration [K$^+$]$_{\rm o}$ (Fig.~\ref{fig:stcoh_exp}A),
as discussed in the previous section.
Spontaneous slow oscillations were recorded at extracellular potassium
concentrations ranging from 1~mM to 7~mM \cite{Sanchez-Vives2000} (see
SI Section S2 for experimental details). These values are on the order of the levels found
typically {\em in vivo}, located around 3~mM \cite{Yamaguchi:1986fk,Amzica:2002uq,Bazhenov:2004fk}.

First we tested whether an increase of the extracellular potassium concentration
led to an increase in the background noise of the network, 
as measured by the standard deviation of the log(MUA) during the DOWN states.
Figure~\ref{fig:stcoh_exp}B confirms that this is indeed the case.
The response of the {population firing rate, measured again in terms of the log(MUA), is shown in Fig.~\ref{fig:stcoh_exp}C} for three different potassium levels. As expected, higher potassium leads to more frequent
UP states. In order to quantify the variability in the duration of the complete UP/DOWN cycle,
we evaluated the coefficient of variation of that quantity as a function of the potassium level.
As shown in Fig.~\ref{fig:stcoh_exp}D, all the experimental trials ($N=13$ slices)
exhibit a minimum in the variability for an intermediate excitability level, although
the minima occur at slightly different extracellular potassium concentrations across slices.
Differences across slices and experiments may arise due to distinct basal excitability
levels secondary to intrinsic connectivity.
To account for these systematic differences, we aligned
the level of potassium to place the minima of all trials at zero, and calculated the average z-score of the CVs. The result, plotted in Fig.~\ref{fig:stcoh_exp}E,
shows a minimum variability for an intermediate level of excitability,
thus confirming the predicted existence of stochastic coherence. Note that the normalization implicit in the z-score computation does not change the relationship between the dependent variables and [K$^+$]$_{\rm o}$.
The results also show that the CV of the UP state is much less
dependent on the excitability level than that of the DOWN state for low excitability (Fig.~\ref{fig:stcoh_exp}F), in qualitative agreement with
the modeling results.
Taken together, our experiments confirm that the network activity 
acts as a collective order parameter that controls the regularity of the global rhythm.

\subsection*{Noise-induced spatial memory}

The neuronal network underlying the emergence of UP/DOWN dynamics in our model is organized in
space: the connection probability between pairs of neurons is higher the closer the neurons are to each other 
(see Online Methods).
This gives the UP states a propagative character: noise-driven initiation of the neuronal activity occurs at a given point in
the network, and propagates away from it with a speed that depends on the strength of the synaptic coupling
and on the excitatory/inhibitory balance
\cite{Compte2003}. We thus ask what is the effect of background synaptic noise,
and of the associated noise-induced regularity, on the spatiotemporal organization of the
UP/DOWN dynamics. In fact, previous theoretical studies \cite{chialvo} have suggested the existence
of noise-induced memory in spatially-extended systems with non-collective excitability, when operating in a regime
of stochastic coherence.

\begin{figure}[htb]
\centerline{
\includegraphics[width=0.28\textwidth]{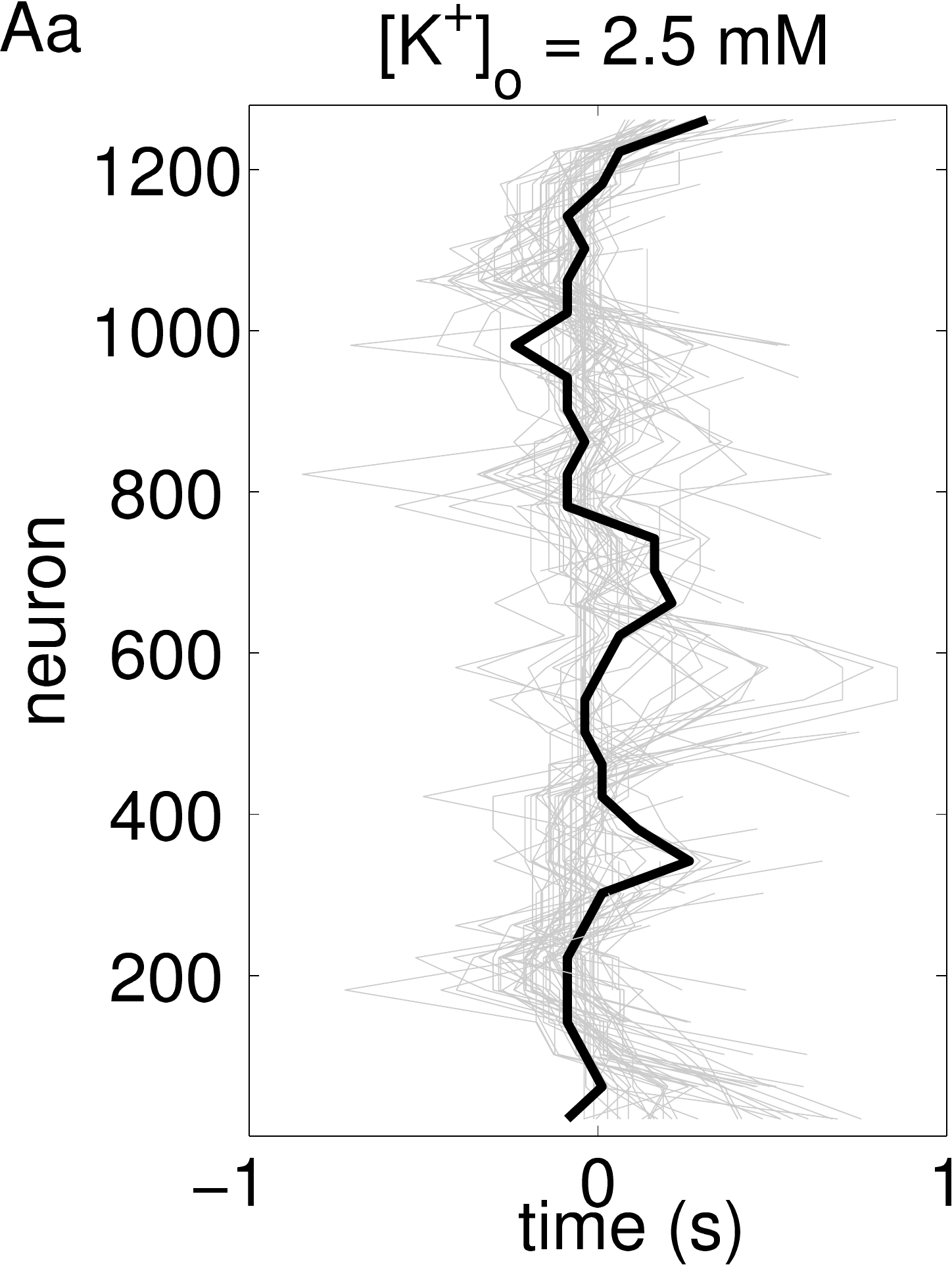}~~~
\includegraphics[width=0.28\textwidth]{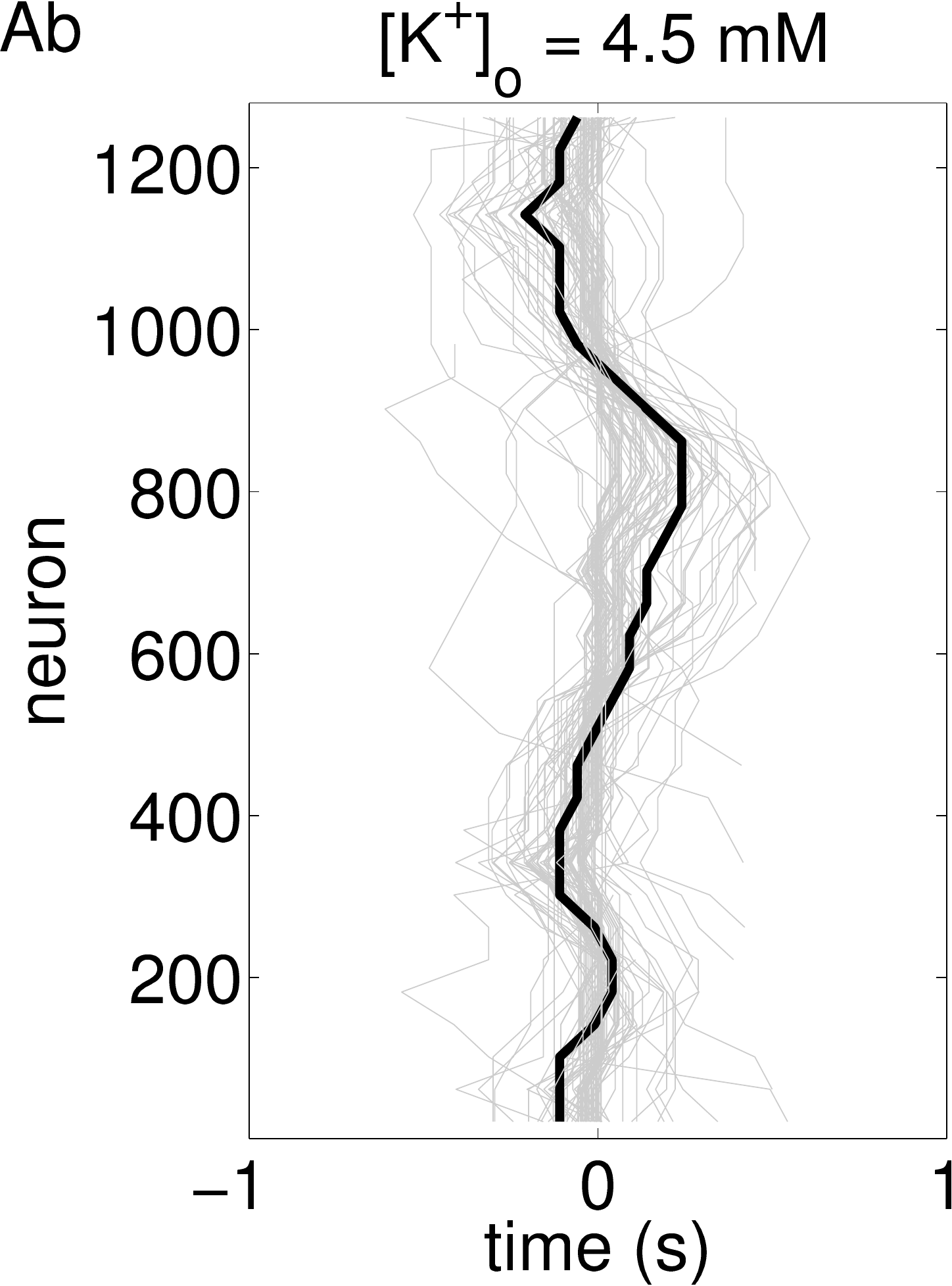}~~~
\includegraphics[width=0.28\textwidth]{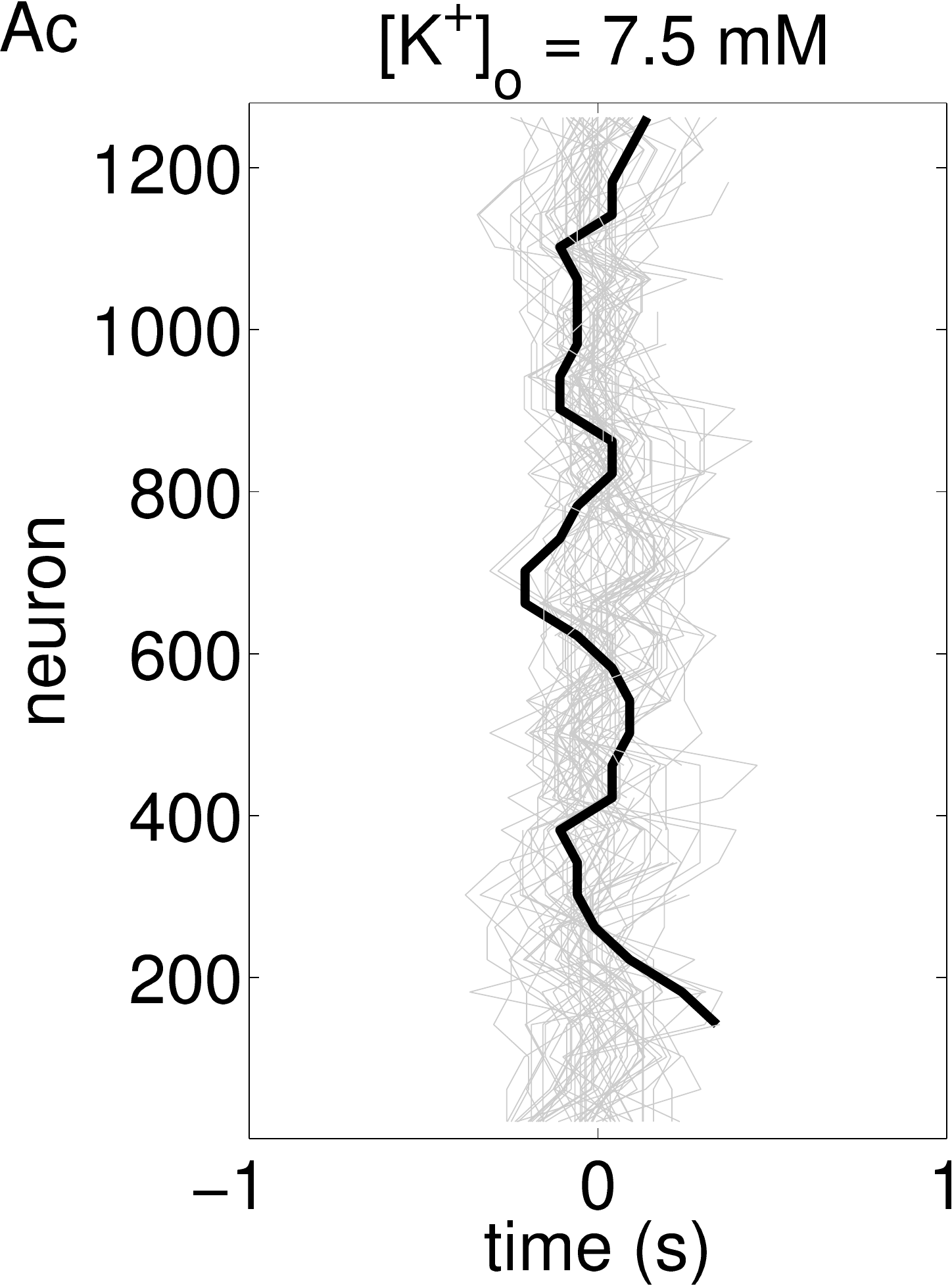}
}
\vskip2mm
\centerline{
\includegraphics[width=0.4\textwidth]{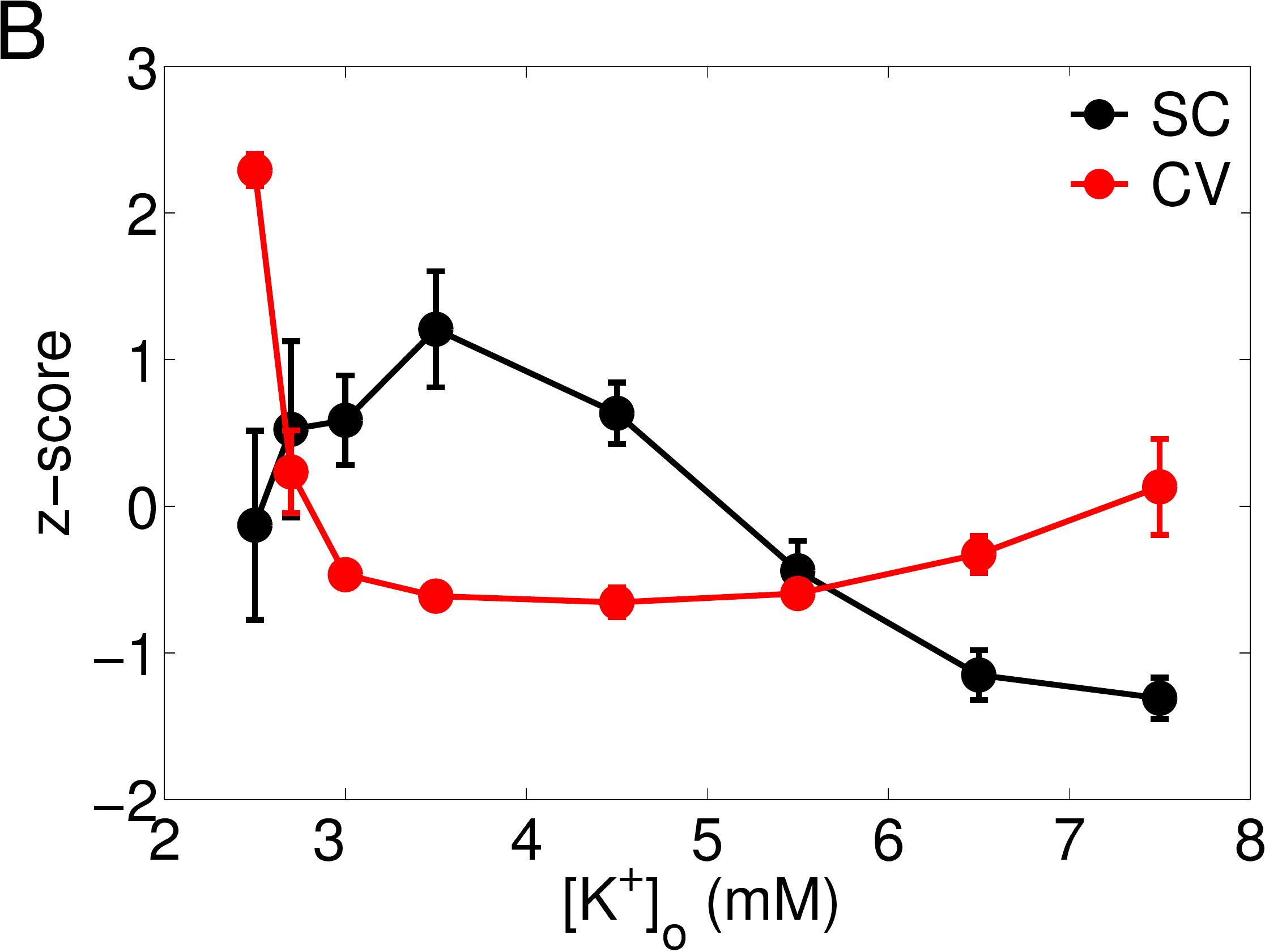}
}
\caption{Noise-induced spatial memory of UP-state initiation in the model. A, {UP wavefronts (grey) centered at the origin as they propagate through} the network, for three different excitability levels, 
controlled by the extracellular potassium concentration [K$^+$]$_{\rm o}$. Specific wavefronts are highlighted in black. B, spatial consistency (SC) of the UP waves and coefficient of variation (CV) of the cycle duration for increasing [K$^+$]$_{\rm o}$. Note that this plot differs from Fig.~\ref{fig:stcoh_mod}C, because the connectivity between excitatory neurons has been decreased in order to better capture the propagation of the UP state through the network.
Error bars in panel B represent corrected sample standard deviation.
\label{fig:memory}}
\end{figure}

To test whether our system can exhibit noise-induced spatial memory, we examined the spatiotemporal
behavior of our neuronal network model for increasing excitability levels. In order to capture clear UP wavefronts we decreased the speed of propagation by reducing the spread of the excitatory connectivity in the model. The results
are presented in Fig.~\ref{fig:memory}A, which shows the wavefronts of the UP states exhibited by the model network (aligned horizontally so that they are superimposed in time) for three different amounts of extracellular potassium
concentration (corresponding to three levels of excitability, and thus of background
synaptic noise).
The wavefronts are shown as contour plots computed over the spike-time histograms of groups of 40
consecutive neurons spanning the entire array.
These plots suggest that the propagation of UP states is rather irregular for both low and high excitability (noise), whereas it follows a well-defined spatial pattern with clear initiation sites for intermediate noise levels. This is quantified in
terms of the spatial consistency (SC) of the UP wavefronts, which is inversely related to the amount of spatial dispersion among the wavefronts shown in Fig.~\ref{fig:memory}A (see Online Methods). This quantity exhibits a clear maximum as the potassium concentration increases, as shown in Fig.~\ref{fig:memory}B. On the other hand, 
the temporal dispersion of the initiation events (as quantified by the CV) shows a clear minimum as the excitability increases. Therefore, our model 
predicts the existence of noise-induced spatial memory in the propagation of UP states
for intermediate levels of background synaptic noise, which concurs with the regularity of the collective dynamics. 

\begin{figure}[htb]
\begin{center}
\begin{minipage}[c]{\textwidth}
\includegraphics[width=0.45\textwidth]{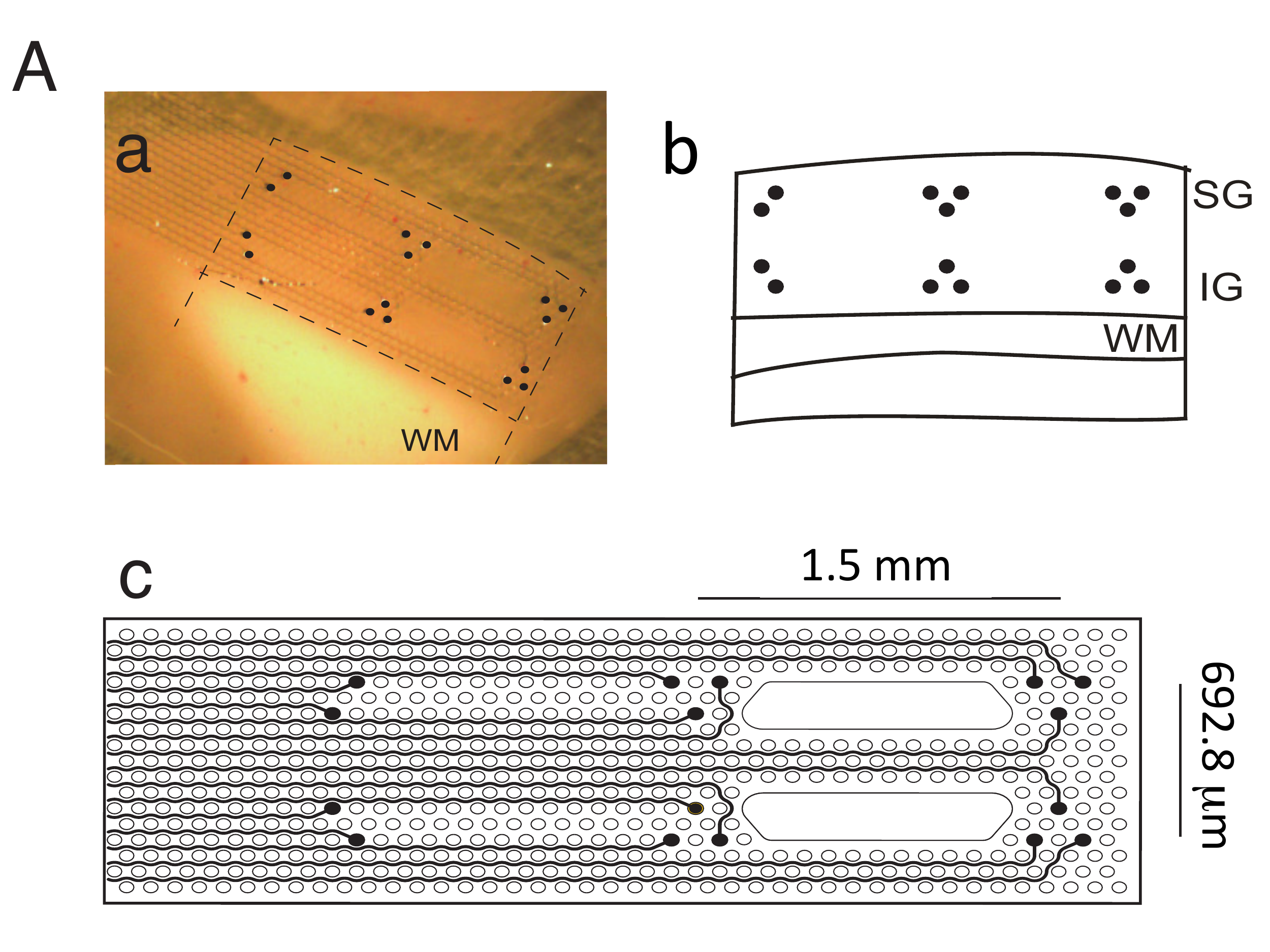}~~
\includegraphics[width=0.45\textwidth]{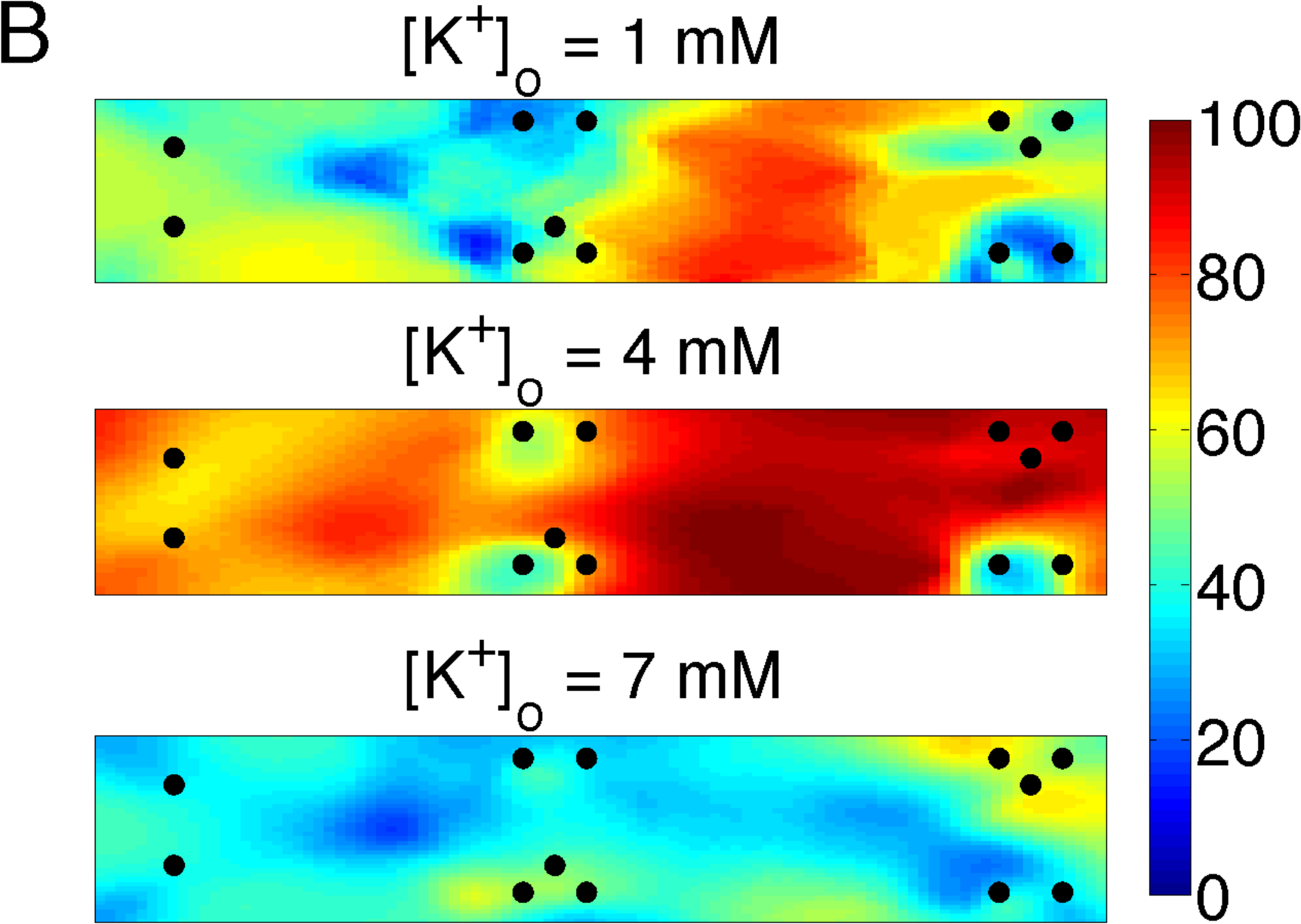}
\end{minipage}
\end{center}
\vskip2mm
\centerline{
\includegraphics[width=0.4\textwidth]{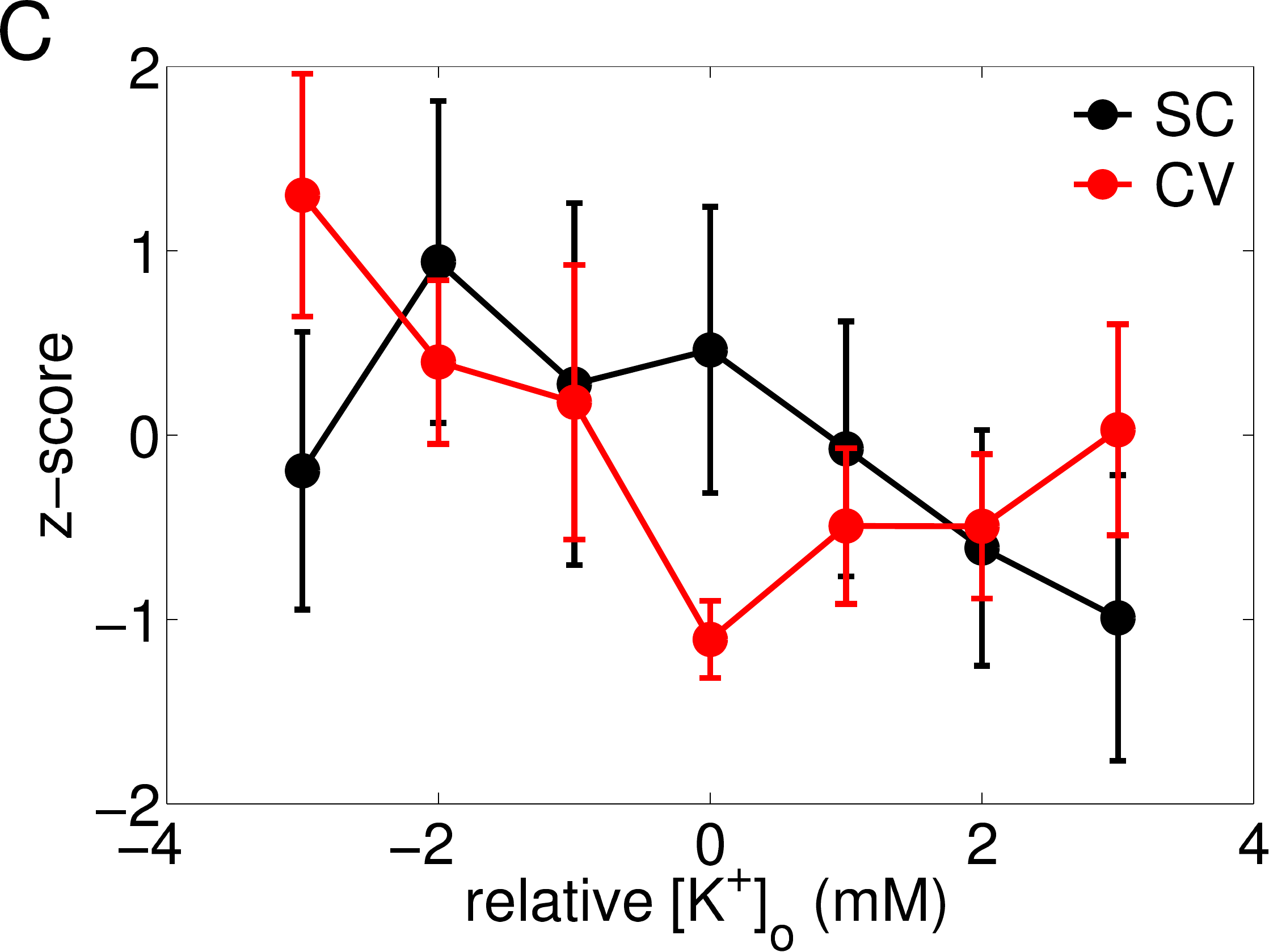}
}

\caption{Noise induced spatial memory of UP-state initiation events in cortical slices.
Aa, illustration of our flexible array with sixteen electrodes (black dots) superimposed on
a visual cortex slice. Ab, scheme of the recording configuration showing the location of the different electrodes
(supragranular layers, SG, and infragranular layers, IG) with respect to the white matter (WM).
Ac, scheme of the two-dimensional electrode array and distances between groups of electrodes. 
B, color coded spatial consistency of the UP wavefronts for
three different levels of extracellular potassium concentration in a typical slice. The black circles indicate the location of
the electrodes.
C, z-score values of the spatial consistency (SC) and temporal coefficient of variation (CV) across the set of experiments (N = 8).
All [K$^+$]$_{\rm o}$ values are referred to the location of the CV minimum for each experiment before averaging. Error bars show the corrected sample standard deviation across experiments.
\label{fig:memory_exp}}
\end{figure}

To verify this theoretical prediction, we measured the electrical activity of cortical slices
with the electrode array shown in Fig.~\ref{fig:memory_exp}A (see SI Section S2). This setup allows us to
monitor the spatiotemporal dynamics of the LFP exhibited by the cortical slice, and we do
so for varying levels of extracellular potassium.
An example of the behavior of the tissue
for a given experimental trial is depicted in Fig.~\ref{fig:memory_exp}B, again for three
different levels of excitability. As we did with the modeling results, we quantified the spatial consistency of the UP wavefronts as described in the Online Methods. This measure is shown in the color maps of Fig.~\ref{fig:memory_exp}B for a given experiment. The plot reveals
that the UP wavefronts follow more regular patterns of propagation (dark red areas) for an intermediate excitability level,
as predicted by our computational model. The behavior of this spatial consistency for increasing extracellular potassium levels is negatively correlated with the temporal dispersion
(coefficient of variation) of the UP/DOWN cycle durations, as illustrated in Fig.~\ref{fig:memory_exp}C, in agreement with the theoretical prediction.
An analysis of the statistical significance of the curves shown in Fig.~\ref{fig:memory_exp}C using mixed-effects modeling
\cite{Baayen:2008fk} allows us to reject the null hypothesis that potassium does not affect neither CV ($p<10^{-10}$) nor SC ($p<0.005$). This supports our conclusion that potassium affects both the temporal
and the spatial dispersion of the UP/DOWN dynamics.

\subsection*{Physiological implications}

It is well known that excitable
systems (of which neurons are an example) can tune noise to enhance the
regularity of their pulsing dynamics \cite{Pikovsky1997,Lindner2004,self} at a characteristic
timescale that is intrinsic to the system \cite{Gang:1993cl}, instead of being imposed externally
\cite{Bulsara1991531,Longtin:1993fk,Douglass:1993ly,Levin:1996hc,Collins:1996kx,Stacey:2000ve,McDonnell:2011rr}. 
This effect has been observed in single neurons \cite{Gu:2002wt} and in
the somatosensory pathway of anesthetized cats \cite{Manjarrez:2002fq}
in response to \emph{ad hoc} noisy inputs. In contrast to these works, here we
have studied a purely cortical {\em in vitro} system under the influence of its own intrinsic noise,
operating in a dynamic regime of slow oscillations that mimic those during slow wave sleep and anesthesia.
Our results reveal that
an intermediate amount of background noise maximize the regularity of the UP/DOWN cycle.
Modulating the regularity of these slow oscillations, which are an emerging property of the network,
could facilitate the control of cortical information processing, by
enabling mechanisms such as communication through coherence \cite{Fries:2015fk,Barardi:2014fk}, or in general any
mechanism that requires a periodic information carrier.
Differently from other studies \cite{Han1999,Horikawa2001} that consider a white noise added to
the membrane voltage, in our model the source of noise arises within the network, in the form of
irregular presynaptic spike
trains affecting all neurons of the network and coming from neighboring areas.
This noise can be tuned experimentally via the
membrane excitability of the individual neurons, determined by the potassium
concentration [K$^+$]$_{\rm o}$ in the extracellular medium.
In this scenario, population activity
appears as a collective phenomenon that underlies the control of rhythmicity.

Both the network model and the experimental recordings confirm that slow UP/DOWN rhythms are
modulated by random fluctuations, 
and achieve a maximum periodicity at an intermediate amplitude of those
fluctuations.
At low excitability levels the probability of eliciting an UP state is low, and the
occurrence of these events fluctuates
strongly on time, giving rise to high variability in the durations of the DOWN states. On the other hand,
for high levels of excitability the refractory period can be overcome more easily (i.e. the system is
re-excited sooner following the UP termination).
This is in agreement
with the behavior observed in
the transition from sleep to awake, or from deep to light anaesthesia \cite{Deco2009},
where the UP/DOWN transitions become more frequent and irregular.
Between the two extremes, the UP states are consistently
evoked at a similar phase of the DOWN state, and a recurrent collective period emerges.
The specifics of this regular UP/DOWN regime depend on the balance
between recurrent excitation and inhibition.

These results shed light on the self-regulation
of cortical dynamics across different states of excitability, and reveal the existence of a
regime of collective stochastic coherence leading to
a maximal regularity of the dynamics for intermediate excitability.    
This is an emerging behavior reminiscent of the phenomenon of coherence
resonance, which has been reported in a large variety of physical and chemical excitable
systems over the years \cite{Lindner2004}. In contrast with those previous studies,
however, the effect described here is an emerging property of the network,
arising from the collective interaction between the neurons. Given the
stochastic and intrinsically emerging
character of brain function, our results might be evidence of the functional 
benefits of noise in the activity of the central nervous system.

\subsection*{Acknowledgements} 
We thank Albert Compte for useful comments, and our colleagues from CSIC-CNM in
Barcelona (X. Villa, R. Villa, G. Gabriel) for providing the recording arrays used in Fig.~\ref{fig:memory_exp}.
This work was supported by the Ministerio de Economia y Competividad and FEDER
(Spain, projects FIS2012-37655-C02-01 --JGO-- and BFU2014-52467-R --MVSV--)
and EU project CORTICONIC (contract number 600806, MVSV).
BR was supported by the FPI programme associated to BFU2011-27094
(Spain, Ministerio de Economia y Competividad). 
JGO acknowledges support from the ICREA Academia programme and from the Generalitat de Catalunya (project 2014SGR0974). 

\subsection*{Author contributions}

BS, MVSV and JGO conceived the research. BS implemented the mathematical model. BS and PB analysed the data. BR performed the experiments. MVSV and JGO supervised the work. BS, MVSV and JGO wrote the manuscript. All authors revised and approved the text.

\section*{Online methods}

\subsection*{Description of the network model}

We consider a network composed of 1280 neurons, 80\% of which are excitatory while the remaining 20\% are inhibitory \cite{Mountcastle1998}. The neurons, modeled as described in Suppl. Section S1, were arranged in two open one-dimensional chains, one for the excitatory population and the other one for the inhibitory population, with one inhibitory neuron placed every four adjacent excitatory neurons. All connections between cells are chemical synapses --no gap junctions are considered--, and each neuron connects with 20$\pm$5 other neurons. The two chains are 5~mm long, with no empty space between adjacent neurons.
We use the coupling architecture introduced by \cite{Compte2003}, in which
the probability that two neurons are connected is determined by the distance {\em x} between them, according to a Gaussian probability distribution $P(x)=\frac{e^{-x^2/2\sigma^2}}{\sqrt{2\pi\sigma^2}}$, with no autapses being allowed. The standard deviation $\sigma$ is set to 250~$\mu$m for the excitatory connections, and 125~$\mu$m for the inhibitory connections. As a reference, the size of an excitatory neuron can be
considered $\sim$5~$\mu$m (50~times smaller than its
$\sigma$) and
that of an inhibitory neuron $\sim$19~$\mu$m (about 13~times smaller than its $\sigma$).
The neurons are also driven by random spike trains drawn from a Poisson distribution, representing
the effect of neurons that are not included explicitly in the network.
Experimentally, this random input corresponds to
fluctuations in the excitability of neurons neighboring the area where the slow oscillation is taking place.

\subsection*{Estimation of the local field potential (LFP)} 

The local field potential (LFP) is computed as the sum of the absolute values of the excitatory and
inhibitory synaptic
currents acting upon the excitatory neurons \cite{Mazzoni2008,Berens:2010fk,Buzsaki2012}:
\begin{equation}
\mathrm{LFP} = R_e \sum\left( | I_{\rm AMPA} |+| I_{\rm NMDA} |+| I_{\rm GABA} | \right)
\label{eq:Appendix_LFPdef}
\end{equation}
The terms $I_{\rm AMPA}$ and $I_{\rm NMDA}$ account for both the external excitatory heterogeneous
Poisson spike train and the recurrent excitatory synaptic current due to network connectivity, respectively.
In turn, $I_{\rm GABA}$ corresponds to
the recurrent inhibitory synaptic current. $R_e$ represents the resistance of a typical electrode used for extracellular measurements, here chosen to be 1~M$\Omega$. {The LFP is sampled at 1~kHz.} The initiation and termination of the UP and DOWN states were identified
by means of the {log(MUA), a measure extracted from the power spectrum of the LFP [Equation~(\ref{eq:Appendix_LFPdef})] within the frequency range 0.2--0.5~kHz (see next section for a complete description)}.

\subsection*{Estimation of the log(MUA)}

The multiunit activity (MUA) is estimated as the power change in the Fourier components of the recorded (or simulated) LFP at high frequencies. The time-dependent MUA is computed from the power spectrum in 50-ms windows, each frequency normalized by the corresponding amplitude of the power spectrum computed over the whole time series, and averaged within the 0.2--1.5~kHz band (0.2--0.5~kHz in the simulations, except in Supp. Fig.~S1B, where the range
0.1--0.5~kHz was used). The MUA signal is then logarithmically scaled and smoothed by a moving average with a sliding window of 80~ms \cite{Reig2010}. The UP and DOWN states were singled out by setting a threshold in the log(MUA) signal. The threshold was set between peaks of the bimodal distributions of log(MUA), corresponding to the UP and DOWN states. The peak related to the DOWN state was used as reference, setting there log(MUA) = 0.
The z-score of the log(MUA) is computed by subtracting from each SD curve its mean value for all
simulations/experiments across the extracellular potassium range, normalizing it by its standard deviation, and averaging the results across replicates.

\subsection*{Spatial consistency analysis}

In order to quantify the regularity of the UP waves we proceed as follows. First, after detection of every UP state, we compute the times at which each electrode crosses a given threshold (as mentioned in the section above). The electrodes whose signal is very noisy are not used for the analysis. On the other hand, we consider only those UP states that are detected in all three columns in which the array is arranged (see Fig.~5Aa,b) --but not necessarily in all the electrodes in each column--. This condition enables us to consider those waves that propagate through all three columns across the slice. We transform those first-passage times into time lags by subtracting the initial time at which the UP wave is first detected somewhere within the array. If one of the propagating UP states is not detected in one of the electrodes used for the analysis, we assign to that electrode for that particular wave a time lag that is an average from other time lags belonging to the 5 most similar propagating wavefronts.

Next we interpolate the time lags using a thin-plate spline by means of the MATLAB function tpaps. The new data points form a grid of 105$\times$49 points. Hence we have as many grids as detected UP states. Each grid is further subdivided into a coarse matrix of 15$\times$7 cells.
We then compute the correlation between each one of these cells with the equivalent cell belonging to the other detected UP waves. Therefore, the correlation between two UP waves leads to 105 ($= 15\times7$) correlation coefficients, and the correlation between $N$ UP waves leads to $\frac{N(N-1)}{2}\times 105$ coefficients.

Finally, we compute the spatial consistency (SC) of the wavefronts in terms of the percentage of highly correlated matrix elements ($>0.7$) between all wave pairs for each of the cells. The color maps in Fig.~5B represent the spatial consistency for all cells in the $15\times7$ coarse matrix.
The same approach is followed in our one-dimensional model, where the time lags are obtained from 16 clusters of consecutive excitatory neurons covering the entire one-dimensional model network, while an interpolation is performed to generate a vector of 49 cells.

\renewcommand{\thesection}{S\arabic{section}}
\renewcommand{\thesubsection}{S\arabic{section}\Alph{subsection}}

\newpage
\centerline{\Large\bf Supplementary Information}

\section{Computational model}
\label{sec:comp}

The dynamical equations for the neuronal membrane voltage and for the ionic channels are based on
those introduced by Compte et al \cite{Compte2003}.
The somatic and dendritic voltages of the excitatory neurons obey the equations:
\begin{align}
C_{\rm m}A_{\rm s}\frac{dV_{\rm s}}{dt}=&-A_{\rm s}\left(I_{\rm L}+I_{\rm Na}+I_{\rm K}+I_{\rm A}+I_{\rm KS}+I_{\rm KNa}\right)- \nonumber \\
& -I_{\rm syn,s}-g_{\rm sd}\left(V_{\rm s}-V_{\rm d}\right) 
\label{eq:Appendix_membraneEs}
\\
C_{\rm m}A_{\rm d}\frac{dV_{\rm d}}{dt}=&-A_{\rm d}\left(I_{\rm Ca}+I_{\rm KCa}+I_{\rm NaP}+I_{\rm AR}\right)- \nonumber \\
& -I_{\rm syn,d}-g_{\rm sd}\left(V_{\rm d}-V_{\rm s}\right),
\label{eq:Appendix_membraneEd}
\end{align}
where $I_{\rm syn,s}$ is the synaptic current coming from the neighboring inhibitory neurons and $I_{\rm syn,d}$ is the synaptic current coming from the neighboring excitatory neurons. 
The two compartments are joined by an electrical coupling of conductance $g_{\rm sd}=1.75\pm0.1~\mu$S such that both voltages are quickly synchronized. Here, the synaptic conductances and $g_{\rm sd}$ are scaled by the surface of the soma, $A_{\rm s}=0.015$~mm$^2$, and the dendrites, $A_{\rm d}=0.035$~mm$^2$. The membrane capacitance is $C_{\rm m}=1~\mu$F/cm$^2$.

The currents through the potassium and sodium channels are,
respectively,
\begin{align}
&I_{\rm K}=g_{\rm K}n^4\left(V-V_{\rm K}\right)\\
&I_{\rm Na}=g_{\rm Na}m^3_{\infty}h\left(V-V_{\rm Na}\right),
\end{align}
which depend on the time-varying probabilities, $x$, that a channel is open:
$$
\frac{dx}{dt}=\phi\left[\alpha_{x}(V)(1-x)-\beta_{x}(V)x\right].
$$
Here $x$ stands for $n$ in the case of the potassium current, and for $m$ and $h$ in the case of
the sodium current. The parameter values used throughout this study are those of Ref.~\cite{Compte2003}:
$g_{\rm K}=10.5$~mS/cm$^2$, $g_{\rm Na}=50$~mS/cm$^2$ and
$g_{\rm L}=0.0667\pm0.0067$~mS/cm$^2$. 
The rate functions $\alpha$ and $\beta$ are:
\begin{align*}
&\alpha_{n}(V)=0.01\frac{V+34}{1-e^{-(V+34)/10}} \\
&\beta_{n}(V)=0.125e^{-(V+44)/25}
\end{align*}
for the gating variable $n$,
\begin{align*}
&\alpha_{m}(V)=0.1\frac{V+33}{1-e^{-(V+33)/10}} \\
&\beta_{m}(V)=4e^{-(V+53.7)/12}
\end{align*}
for the gating variable $m$, and
\begin{align*}
&\alpha_{h}(V)=0.07e^{-(V+50)/10} \\
&\beta_{h}(V)=\frac{1.0}{1+e^{-(V+20)/10}}
\end{align*}
for the gating variable $h$.                                                                                                                                                                                                                                                                                                                                                                                                                                                                                                                                                                                                                                                                                                                                                                                                                                                                                                                                                                                                                                                                                                                                                                                                                                                                                                                                                                                                                                                                                                                                                                                                                                                                                                                                                                                                                                                                                                                                                                                                                                                                                                                                                                                                                                                                                                                                                                                                                                                                                                                                                                                                                                                                                                                                                                                                     
Due to the rapid activation of $m$ we replace it by its steady-state
value $m_{\infty}=\frac{\alpha_{m}}{\alpha_{m}+\beta_{m}}$. The leak current is a passive channel
represented by $I_{\rm L}=g_{\rm L}\left(V-V_{\rm L}\right)$.

The time-varying probabilities $x$ that a channel is open can also be expressed as:
$$
\frac{dx}{dt}=\frac{1}{\tau_x(V)}\phi\left[x_{\infty}(V)-x\right].
$$
The rest of ionic channels are described in terms of $x_{\infty}(V)$ and $\tau_x(V)$, and the temperature factor $\phi$ is set to 1.
For the fast A-type K$^+$-channel, $I_{\rm A}=g_{\rm A}m^3_{\infty}h\left(V-V_{\rm K}\right)$, with:
\begin{align*}
&m_{\infty}(V)=\frac{1}{1+e^{-(V+50)/20}} \\
&h_{\infty}(V)=\frac{1}{1+e^{(V+80)/6}} 
\end{align*}
where $\tau_h=15$~ms and $g_{\rm A}=1$~mS/cm$^2$. For the non-inactivating
K$^+$-channel, $I_{\rm KS}=g_{\rm KS}m\left(V-V_{\rm K}\right)$, with:
\begin{align*}
&m_{\infty}(V)=\frac{1}{1+e^{-(V+34)/6.5}} \\
&\tau_m(V)=\frac{8}{e^{-(V+55)/30}+e^{(V+55)/30}}
\end{align*}
where $g_{\rm KS}=0.576$~mS/cm$^2$. For the persistent sodium channel,
$I_{\rm NaP}=g_{\rm NaP}m^3_{\infty}\left(V-V_{\rm Na}\right)$, with:
$$
m_{\infty}(V)=\frac{1}{1+e^{-(V+55.7)/7.7}},
$$
where $g_{\rm NaP}=0.0686$~mS/cm$^2$. For the inward rectifying K$^+$-channel,
$I_{\rm AR}=g_{\rm AR}h_{\infty}\left(V-V_{\rm K}\right)$, with:
$$
h_{\infty}(V)=\frac{1}{1+e^{(V+75)/4}}
$$
where $g_{\rm AR}=0.0257$~mS/cm$^2$. For the high-threshold Ca$^{2+}$-channel, $I_{\rm Ca}=g_{\rm Ca}m^2_{\infty}\left(V-V_{\rm Ca}\right)$:
$$
m_{\infty}(V)=\frac{1}{1+e^{-(V+20)/9}} 
$$
where $g_{\rm Ca}=0.43$~mS/cm$^2$.

Two more channels are considered, which depend on ionic concentrations: the Ca$^{2+}$-dependent K$^+$-channel and the Na$^+$-dependent K$^+$-channel:
\begin{align*}
&I_{\rm KCa}=g_{\rm KCa}\frac{[{\rm Ca}^{2+}]}{[{\rm Ca}^{2+}]+K_D}\left(V-V_{\rm K}\right){\rm,\qquad and}\\
&I_{\rm KNa}=g_{\rm KNa}w_{\infty}([{\rm Na}^{+}])\left(V-V_{\rm K}\right) 
\end{align*}
\noindent respectively, where $K_D=30$~mM, $g_{\rm KCa}=0.57$~mS/cm$^2$, $w_{\infty}([{\rm Na}^{+}])=0.37/\left(1+\left(38.7/[{\rm Na}^+]\right)^{3.5}\right)$ and $g_{\rm KNa}=1.33$~mS/cm$^2$. Moreover, the intracellular concentration of calcium follows the dynamical equation:
$$
\frac{d[{\rm Ca}^{2+}]}{dt}=-\alpha_{\rm Ca}A_{\rm d}I_{\rm Ca}-\frac{[{\rm Ca}^{2+}]}{\tau_{\rm Ca}},
$$
where $\alpha_{\rm Ca}$=5~mM/(mA$\cdot$ms) and $\tau_{\rm Ca}=15$~ms. The intracellular sodium concentration obeys the equation:
\begin{align*}
&\frac{d[{\rm Na}^+]}{dt}=-\alpha_{\rm Na}\left(A_{\rm s}I_{\rm Na}+A_{\rm d}I_{\rm NaP}\right)- \nonumber \\
&\quad -\left(R_{\rm pump}+D\xi_{\rm OU}(t)\right)\left(\frac{[{\rm Na}^+]^3}{[{\rm Na}^+]^3+15^3}-\frac{[{\rm Na}^+]^3_{\rm eq}}{[{\rm Na}^+]^3_{\rm eq}+15^3}\right)
\end{align*}
\noindent where $\alpha_{\rm Na}=10\pm2$~mM/(mA$\cdot$ms), $R_{\rm pump}=0.008\pm0.0018$~mM/ms and $[{\rm Na}^+]_{\rm eq}=9.5$~mM. Note that the sodium dynamics obeys the stoichiometry of
the Na$^+$-K$^+$ pump, which releases three sodium ions for each potassium ion brought inside the neuron.
The reversal potentials of the different channels are $V_{\rm K}=-100$~mV, $V_{\rm Na}=55$~mV,
$V_{\rm Ca}=120$~mV and $V_{\rm L}=-60.95\pm0.3$~mV.

For the inhibitory neurons only the somatic compartment is modeled, with its membrane
voltage following the dynamical equation:
\begin{equation}
C_{\rm m}A_{\rm i}\frac{dV}{dt}=-A_{\rm i}\left(I_{\rm L}+I_{\rm Na}+I_{\rm K}\right)-I_{\rm SYN,i},
\label{eq:Appendix_membraneI}
\end{equation}
where $A_{\rm i}=0.02$~mm$^2$. $I_{\rm SYN,i}$ stands for the sum of both the excitatory and inhibitory currents coming from the presynaptic neurons. Contrary to the excitatory neurons, the inhibitory ones are described with the minimum neuronal model for action potential generation. The sodium, potassium and leak current follow the same formalism as in the above excitatory neuronal model with the following $\alpha$ and $\beta$ rate functions:
\begin{align*}
&\alpha_{n}(V)=0.05\frac{V+34}{1-e^{-(V+34)/10}} \\
&\beta_{n}(V)=0.625e^{-(V+44)/80}
\end{align*}
for the gating variable $n$,
\begin{align*}
&\alpha_{m}(V)=0.5\frac{V+35}{1-e^{-(V+35)/10}} \\
&\beta_{m}(V)=20e^{-(V+60)/18}
\end{align*}
for the gating variable $m$, and
\begin{align*}
&\alpha_{h}(V)=0.35e^{-(V+58)/20} \\
&\beta_{h}(V)=\frac{5.0}{1+e^{-(V+28)/10}}
\end{align*}
for the gating variable $h$.                                      

The temperature factor $\phi$ is set to 1. The maximal conductances are $g_{\rm Na}=35$~mS/cm$^2$, $g_{\rm K}=9$~mS/cm$^2$ and $g_{\rm L}=0.1025\pm0.0025$~mS/cm$^2$, and reversal potentials $V_{\rm K}=-90$~mV, $V_{\rm Na}=55$~mV and $V_{\rm L}=-65\pm0.15$~mV. All parameters are kept constant except in the cases given as mean $\pm$ SD. SD corresponds to the standard deviation of a Gaussian distributed parameter over the population. 

The Nernst equation establishes the relation between the reversal potential of a ionic channel $V_{\rm ion}$ (or equilibrium potential at which there is no net flow of ions across the channel) and the ratio $\frac{C_{\rm o}}{C_{\rm i}}$ between the external and internal concentration of the corresponding ion:
\begin{eqnarray}
V_{\rm ion}=\frac{RT}{zF}\ln\left(\frac{C_{\rm o}}{C_{\rm i}}\right)
\label{eq:eqNernst}
\end{eqnarray}  
\noindent where {\em R} is the universal gas constant (8.314~J/Kmol), {\em T} is the temperature in Kelvin, {\em z} is the number of elementary charges of the ion and {\em F} is the Faraday constant (96.845~C/mol).
For the potassium ion ($z=1$) we have set $C_{\rm i}$ (or [K$^+$]$_{\rm i}$) to 150~mM and varied $C_{\rm o}$ (or [K$^+$]$_{\rm o}$) from 2.5~mM ($V_{\rm K}\sim$ -108~mV)  to 7.5~mM ($V_{\rm K}\sim$ -79~mV).

We consider two types of excitatory synapses, mediated by AMPA and NMDA \cite{Wang1999}:
\begin{align}
&I_{\rm AMPA}=g_{\rm AMPA}s(t)(V-E_{\rm syn}) 
\label{eq:Appendix_AMPA}
\\
&I_{\rm NMDA}=g_{\rm NMDA}s(t)(V-E_{\rm syn})\frac{1}{1+[{\rm Mg}^{2+}]e^{-0.062V/3.57}}
\label{eq:Appendix_NMDA}
\end{align} 
where the reversal synaptic potential is $E_{\rm syn}=0$~mV and the extracellular magnesium concentration is $[{\rm Mg}^{2+}]=1.0$~mM. The $g_{\rm AMPA}$ and $g_{\rm NMDA}$ peak conductances are given in Table~1.

The time-dependent part of the corresponding conductances, {\em s(t)},
follows the first-order kinetics:
\begin{equation}
\frac{ds}{dt}=\phi\left(\alpha_sx(1-s)-s/\tau_s\right),
\end{equation} 
with
\begin{equation}
\frac{dx}{dt}=\phi\left(\alpha_x\sum\limits_{j}\delta(t-t_j)-x/\tau_x\right)
\end{equation} 
where $\phi=1$ and the temporal time constants are shown in Table~2.

The inhibitory neurons are mediated by GABA, with corresponding synaptic current equal to
$I_{\rm GABA}=g_{\rm GABA}s(t)(V-E_{\rm syn})$, where $E_{\rm syn}=-70$~mV. The $g_{\rm GABA}$ peak conductance is given in Table~1 and {\em s(t)} again follows first-order kinetics:
\begin{eqnarray}
\frac{ds}{dt}&=&\alpha_i\sum\limits_{j}\delta(t-t_j)(1-s)-\frac{1}{\tau_i}s\,,
\end{eqnarray} 
where $\alpha_i=0.9$~ms$^{-1}$ and $\tau_i=10$~ms.

\renewcommand{\thetable}{S\arabic{table}}
\begin{table}[htb]
\begin{tabular*}{\hsize}{@{\extracolsep{\fill}}lcc}
{Synapse} & {Conductance on inh.} & {Conductance on exc.}\cr
 \hline
{\rm GABA} & $240$~nS & $480$~nS \cr
{Recurrent AMPA} & $2.25$~nS & $7.35$~nS \cr
{Recurrent NMDA} & $3.0$~nS & $8.0$~nS\cr
{External AMPA} & $2.25$~nS & $7.35$~nS\cr
{External NMDA} & $2.0$~nS & $8.0$~nS \cr
\hline
\end{tabular*}
\caption{Synaptic conductances for all the possible connections}
\end{table}

\begin{table}[htb]
\begin{tabular*}{\hsize}{@{\extracolsep{\fill}}lcccc}
{Synapse} & $\alpha_x$ & $\tau_x$ & $\alpha_s$ & $\tau_s$ \cr
\hline
{\rm AMPA}  & $1.0$ & $0.05$~ms & $1.0$~ms${^{-1}}$ & $2.0$~ms  \cr
{\rm NMDA}  & $1.0$ & $2.0$~ms & $1.0$~ms$^{-1}$ & $80$~ms \cr
\hline
\end{tabular*}
\caption{Synaptic time constants for AMPA and NMDA synapses}
\end{table}

Additionally, all neurons receive an heterogeneous Poisson train of excitatory presynaptic potentials
with an event rate (mean = 50~spikes/s) that varies following an Ornstein-Uhlenbeck process {$\xi_{\rm OU}(t)$
of amplitude $D=500$~spikes/s}, which mimics the synaptic input coming from the rest of the cortical slice
that is not being directly simulated by the set of Equations~(\ref{eq:Appendix_membraneEs})-(\ref{eq:Appendix_membraneEd}) and (\ref{eq:Appendix_membraneI}). A noisy realization is obtained following \cite{toralbook}: 
$$
\xi_{\rm OU}(t+h)=\xi_{\rm OU}(t)e^{-h/\tau}+\sqrt{D\frac{\left(1-e^{-2h/\tau}\right)}{2\tau}}u(t+h),
$$
where $\xi_{\rm OU}(0)=\sqrt{\frac{1}{2\tau}}u(0)$, and $u(t)$ is a Gaussian white noise
and correlation time $\tau=16$~ms.

The model has been integrated using the Heun algorithm \cite{toralbook}, with a
time step of 0.05~ms.

\section{Experimental methods}

{\subsection*{Slice preparation and maintenance}}

Four- to eight-month-old ferrets were anesthetized and quickly decapitated according to the European Union guidelines (Strasbourg 3/18/1986), as approved by the local ethical committee. The brain was quickly removed and placed in oxygenated cold solution ($4-10^{\rm o}$C)  in which NaCl was replaced with sucrose \cite{Sanchez-Vives2012}. Coronal visual slices (400$\mu$m) were cut in a vibratome. After preparation, slices were transferred to an interface style recording chamber (Fine Sciences Tools, Foster City, CA) and superfused with an equal mixture in volume of the normal artificial cerebrospinal fluid (ACSF) and the sucrose-substituted solution during 15~min. Next, during 1.5~h a normal ACSF was applied containing (in mM): NaCl, 126; KCl, 2.5; MgSO4, 2; Na2HPO4, 1; CaCl2, 2; NaHCO3, 26; dextrose, 10, and aerated with 95\% O2, 5\% CO2 to a final pH of 7.4. Following this, a modified slice solution was superfused during $1-2$~h, having the same ionic composition except for different levels of (in mM): KCl, 4; MgSO4, 1 and CaCl2, 1. Finally, in order to tune the excitability of the network, KCl was applied to the bath in increasing concentrations from 1 to 8~mM. Measurements were taken after $15-20$~min (the time it takes to get a stable concentration in the bath of the interface chamber used). The bath temperature was set at $34-36^{\rm o}$C.

\subsection*{Extracellular local field potential (LFP) recordings}

{LFP recordings were obtained with $2-4$~M$\Omega$ tungsten electrodes (FHC, Bowdoinham, ME) or a multichannel array consisting of 16 gold electrodes plated with platinum black, disposed in a recording grid as shown} in Figure 5A in the main text.  A 16-electrode microprobe including
an array of holes was designed and fabricated using SU-8 negative photoresist. In particular, the design of the
flexible microprobe consisted on six groups of electrodes placed in-between the holes. In each of the recording
points we had 2-3 closely positioned electrodes (separated by 200 $\mu$m). The electrodes were positioned in
supra and infragranular layers (692 $\mu$m apart) in 3 different cortical columns (1500 $\mu$m apart).
The electrodes
were 50 $\mu$m in diameter, which resulted in high impedance values (at 1~kHz, $| Z | \sim 10$~M$\Omega$).
Thus, to enhance the electrode behavior, decreasing of the impedance value was required. To that end, the effective
surface area of the electrode was increased by electrochemically coating of a high-roughness layer of black platinum.
This coating allowed for a 2-fold decrease in the impedance values.

Neural activity was referenced to an electrode
placed at the bottom of the chamber in direct contact with the artificial cerebrospinal fluid (ACSF). Electrode
impedances and phases were tested
with known signals prior to the recordings for each recording grid, to exclude the possibility of phase delays or distortion.
Signals were acquired unfiltered at 10 KHz with a Multichannel System amplifier, and digitized with a 1401 interface
and Spike2 software (CED, Cambridge, UK). No filters were added during the recording stage to avoid signal distortion. 
{The LFP is sampled at 5~kHz}.

\section{Response to external variability}

As we have seen in Fig.~1 of the main text, increasing excitability leads to a rise not
only in noise (Fig.~1C) but also in firing rate (Fig.~1B).
In order to discern which one of these two
features gives rise to the minimum in CV shown in Fig.~2C, we evaluate
the response of the network to increasing external variability
while [K$^+$]$_{\rm o}$ is kept constant at 2.5~mM. 
Our computational model allows us to directly tune the synaptic noise through the stochastic external input, something that cannot be done experimentally. Numerical simulations of the network for increasing external variability, presented in Suppl. Fig.~\ref{fig:stcoh_mod_control}, show that the behavior obtained in Fig.~2C also holds when noise is directly increased. As shown in that figure, higher noisy inputs have a negligible effect on the neuronal firing rate during the UP states (Fig.~\ref{fig:stcoh_mod_control}Bc, to be compared with Fig.~\ref{fig:stcoh_mod_control}Ac), while the variability in the UP/DOWN oscillation period still shows a minimum (Fig.~\ref{fig:stcoh_mod_control}Bb, to be compared with Fig.~\ref{fig:stcoh_mod_control}Ab). Thus, our computational model
confirms that an intermediate level of background noise maximizes the regularity of UP/DOWN oscillations
in networks of cortical neurons, in a manner that is not associated with the increasing in firing rate.

\renewcommand{\thefigure}{S\arabic{figure}}
\setcounter{figure}{0}

\begin{figure}[htb]
\centerline{
\includegraphics[width=0.28\textwidth]{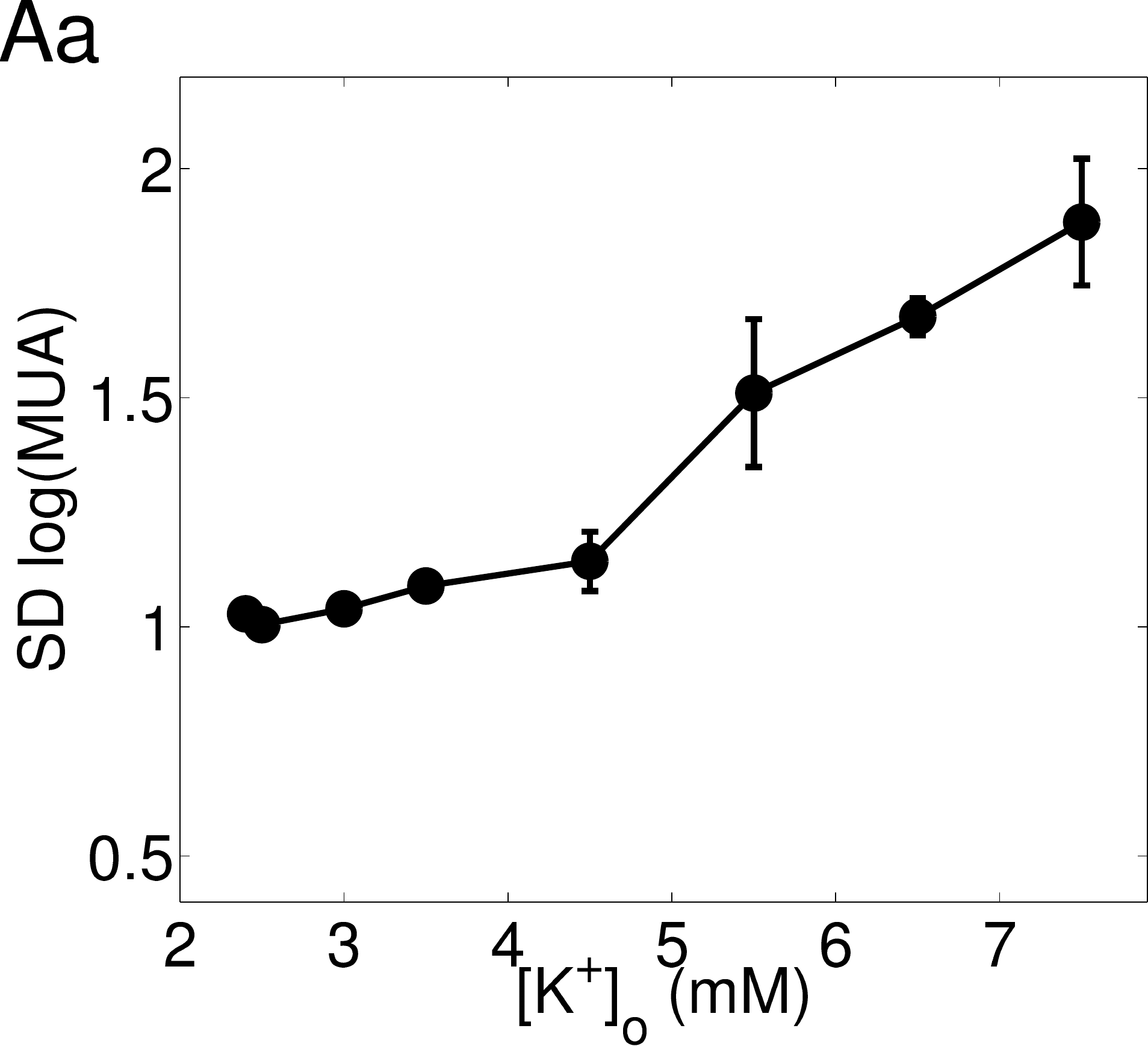}~~~
\includegraphics[width=0.28\textwidth]{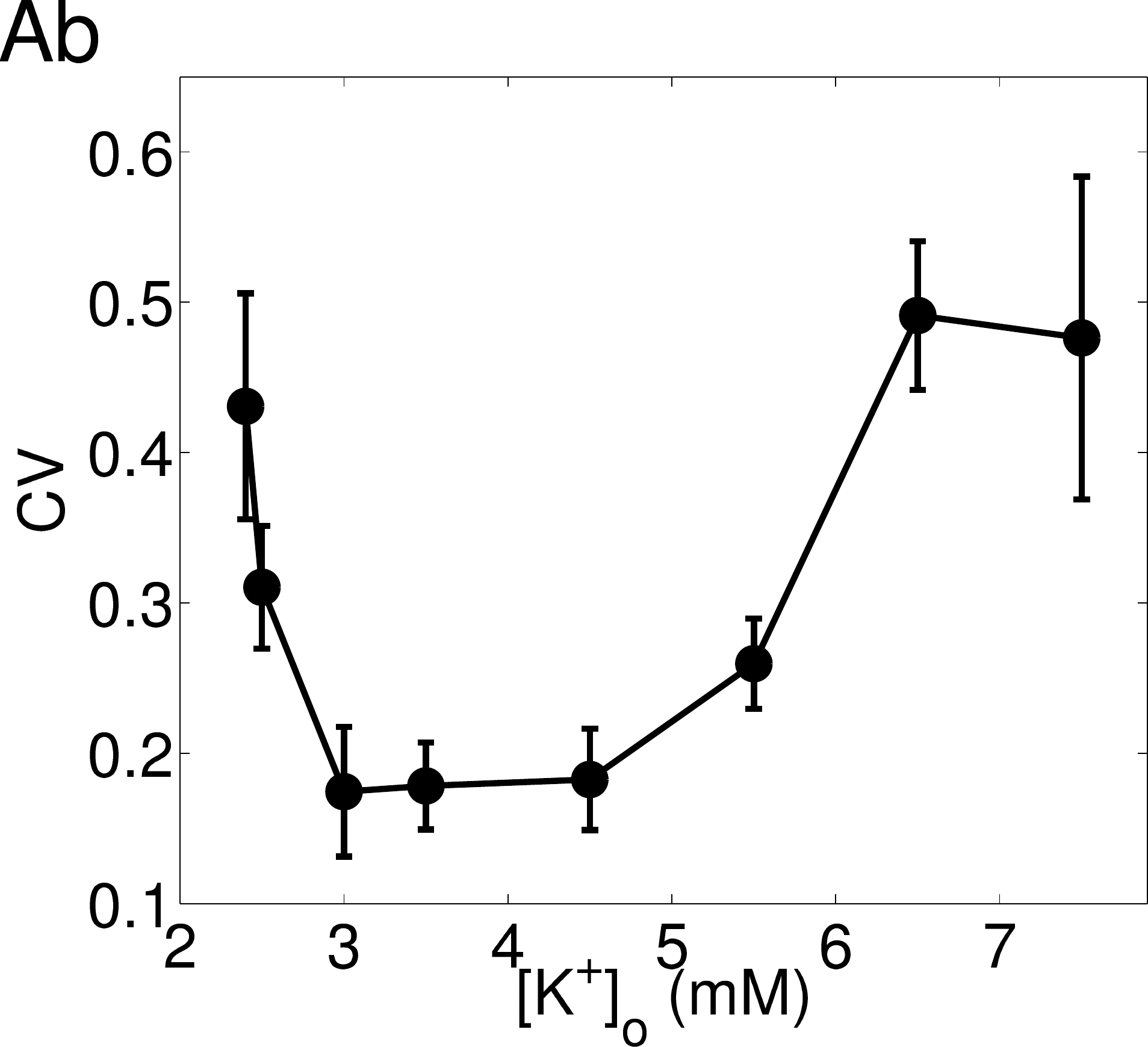}~~~
\includegraphics[width=0.28\textwidth]{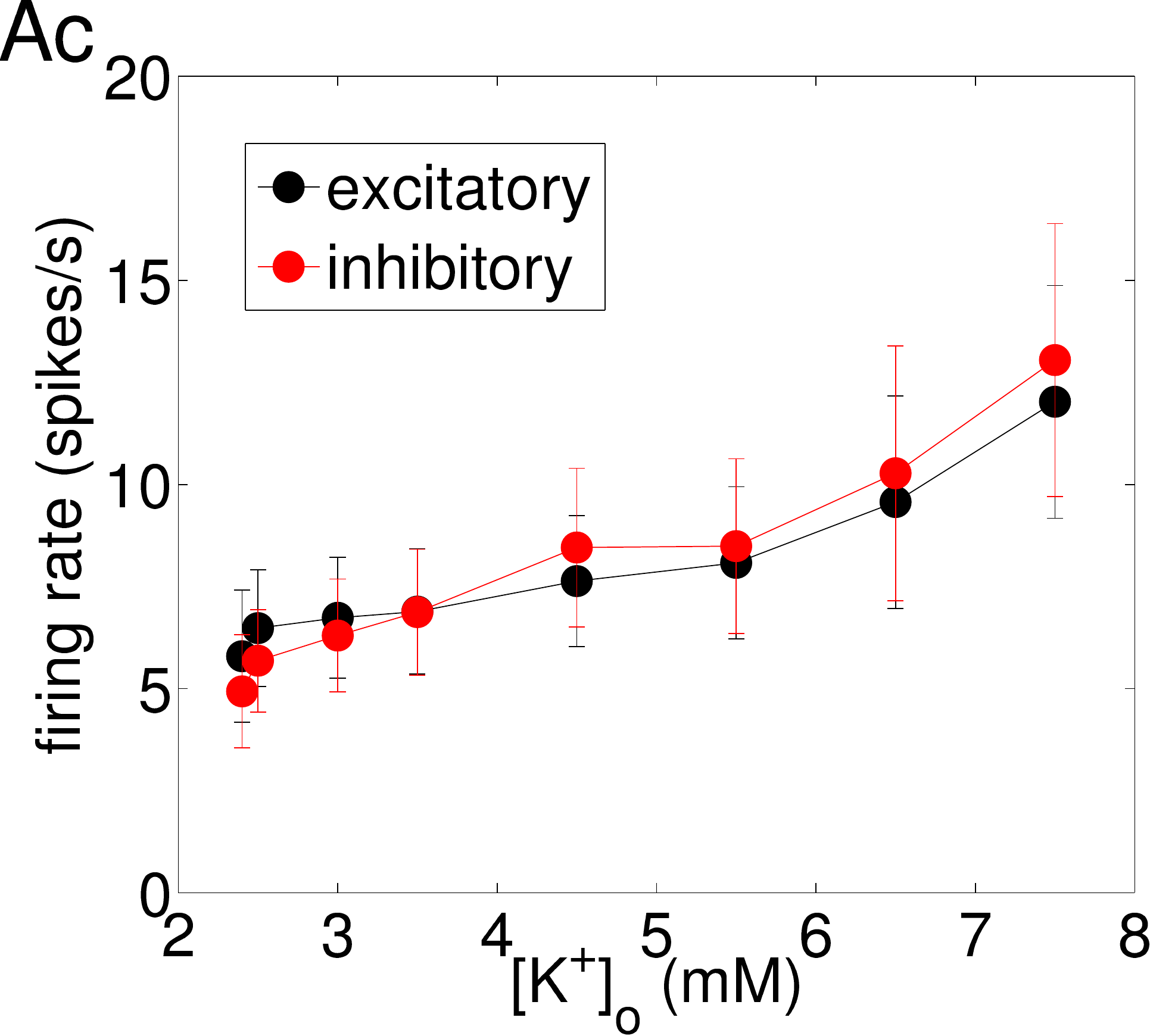}
}
\vskip2mm
\centerline{
\includegraphics[width=0.28\textwidth]{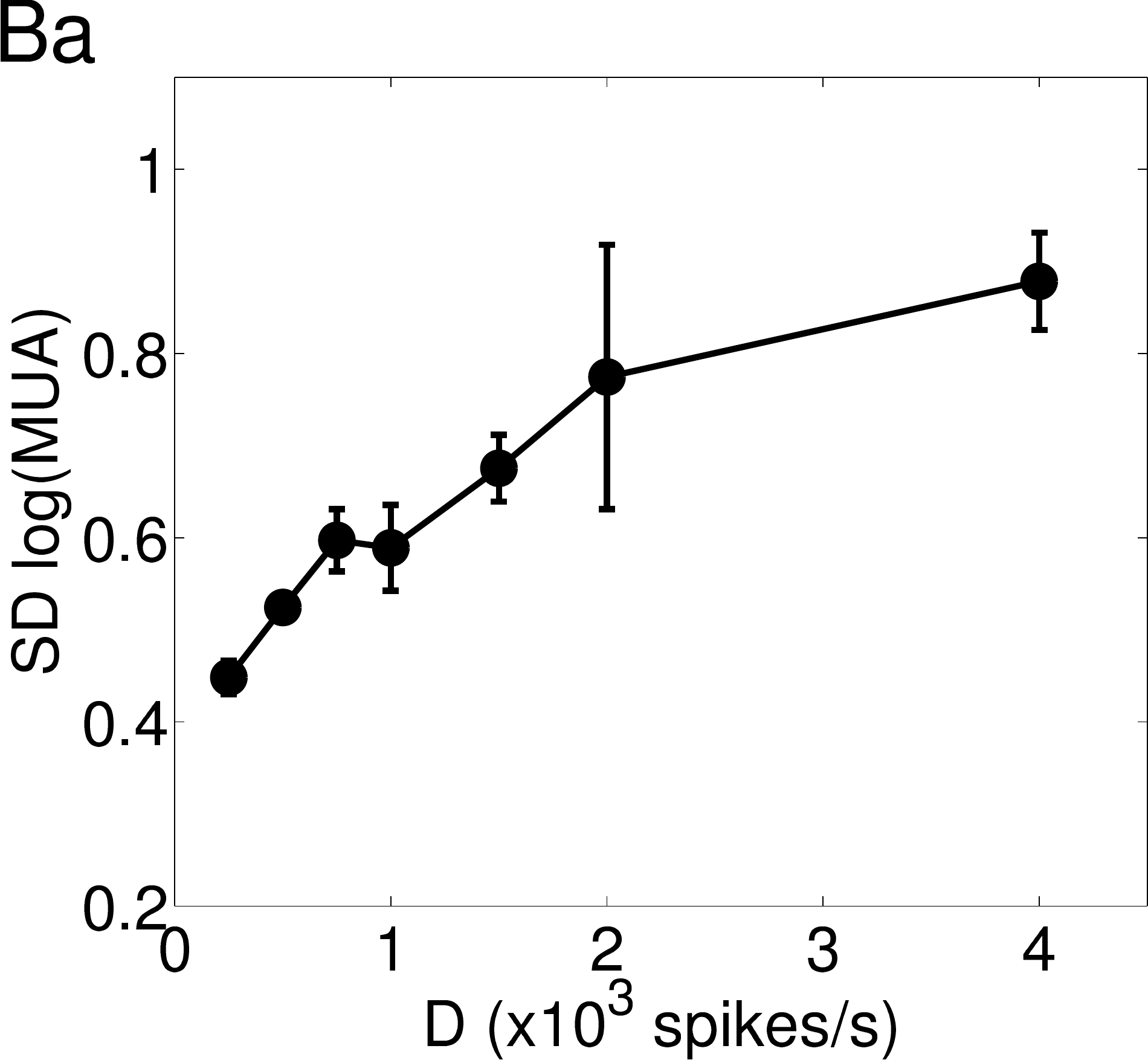}~~~
\includegraphics[width=0.28\textwidth]{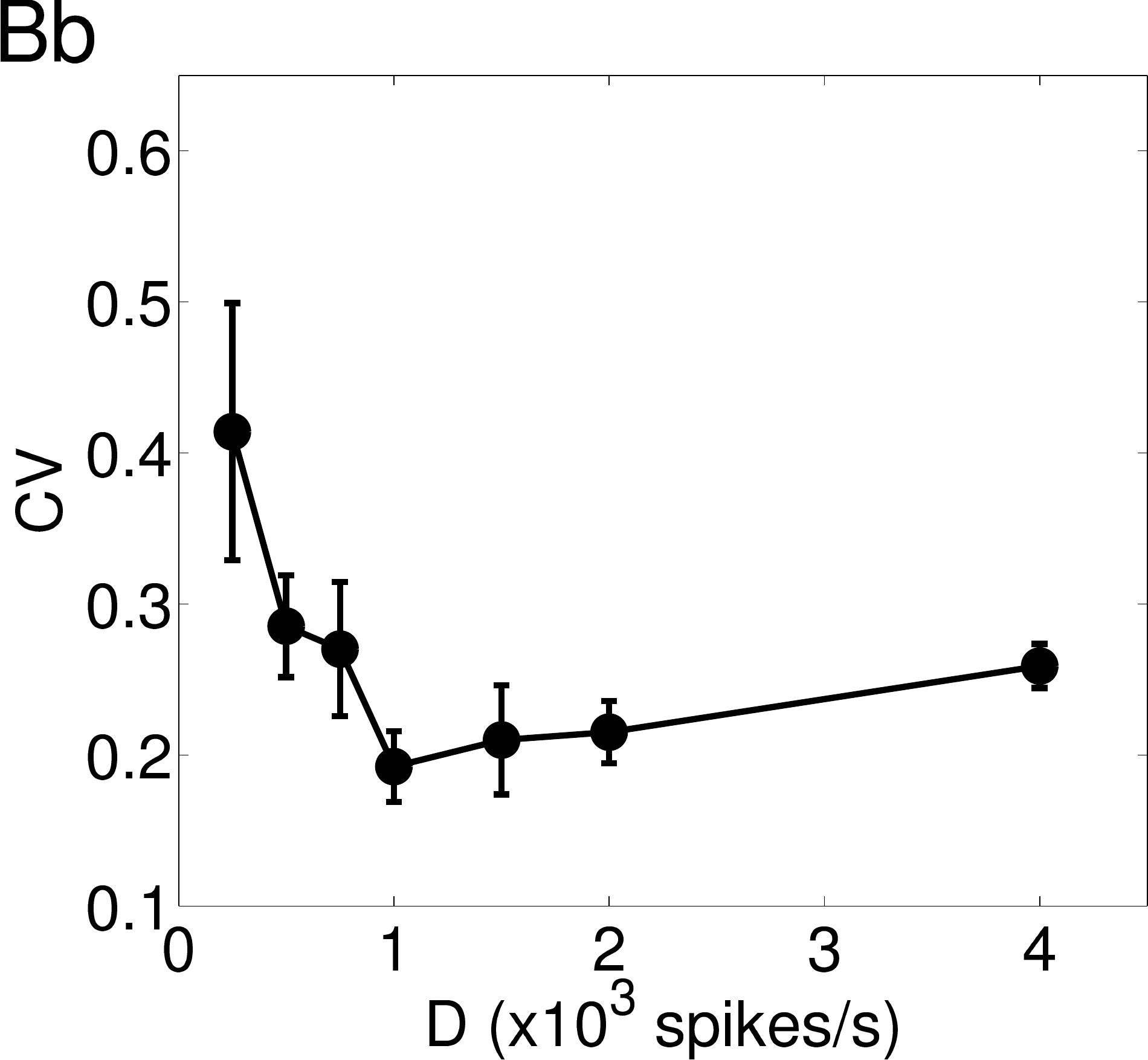}~~~
\includegraphics[width=0.28\textwidth]{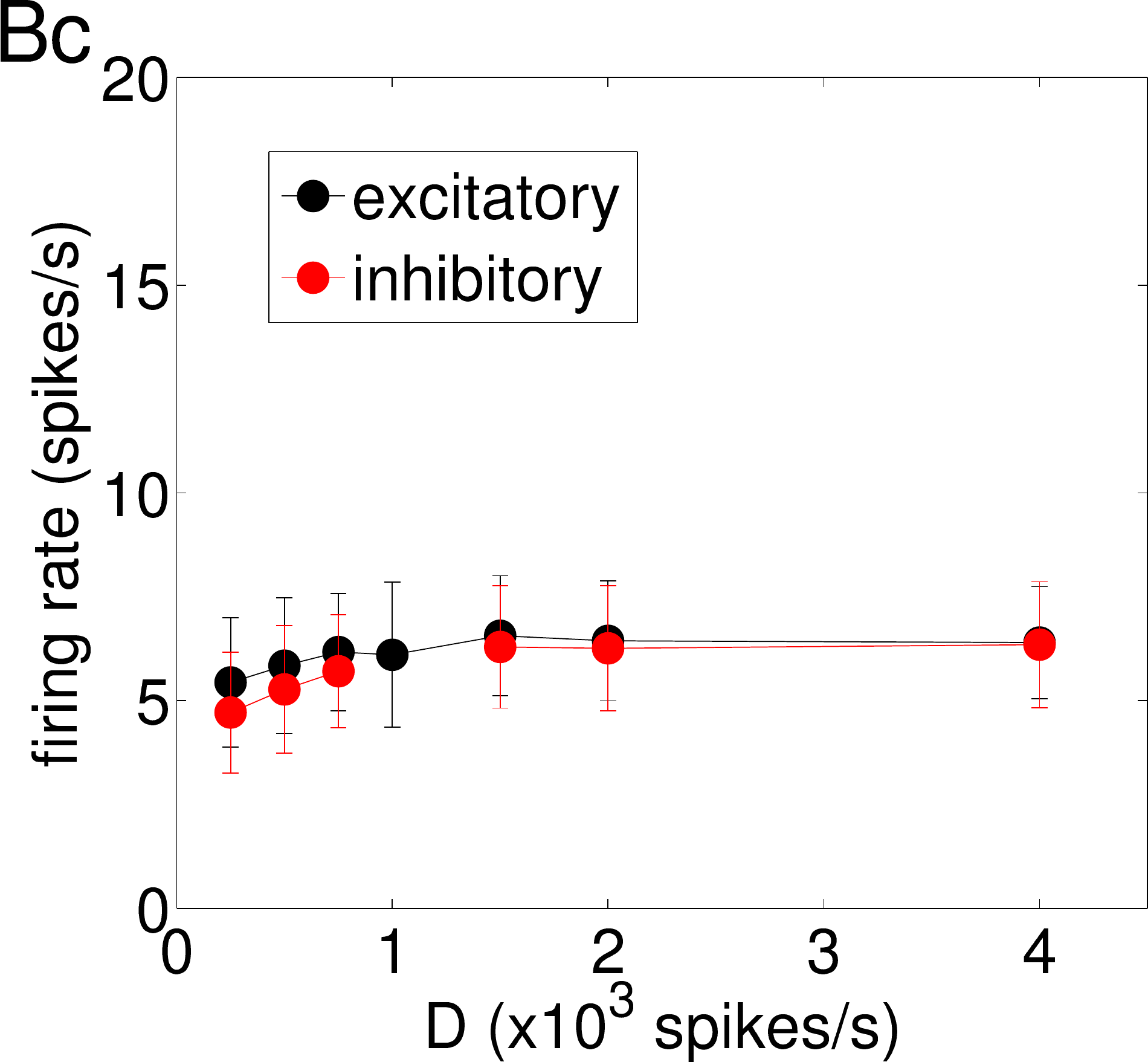}
}
\caption{Effect of increasing {the noise in the computational network in two different ways: A (top row), by increasing extracellular potassium concentration ([K$^+$]$_{\rm o}$); and B (bottom row), by adding random fluctuations to the instantaneous rate of the external spike train, in the form of an Ornstein-Uhlenbeck stochastic process of increasing amplitude $D$,
without varying the intrinsic excitability of the network (see description of the model in section \ref{sec:comp} above).
The left-column plots (a) show the standard deviation of the log(MUA) signal during the DOWN
states, the middle-column plots (b) show the CV of the duration of the full UP/DOWN cycle, and the right-column plots (c)
display the
corresponding population-averaged firing rates for both excitatory (black) and inhibitory (red) neurons. Note that the (a,b) plots
display the absolute values (not z-scores) of the variables, in contrast with the figures shown in the main text. Plots A correspond to the results shown in the main text, plots B correspond to averages over 5 realizations of the numerical simulations.} 
\label{fig:stcoh_mod_control}}
\end{figure}


\end{document}